# A Catalog of Stellar and Dust Properties for 500,000 Stars in the Southwest Bar of the Small Magellanic Cloud

Petia Yanchulova Merica-Jones[1], Karl Gordon[1,2], Karin Sandstrom[3], Claire E. Murray[1,4], L. Clifton Johnson[5], Julianne J. Dalcanton[6,7], Julia Roman-Duval[1], Jeremy Chastenet[2], Benjamin F. Williams[6], Daniel R. Weisz[8], and Andrew E. Dolphin[9,10]

[1] Space Telescope Science Institute, 3700 San Martin Drive, Baltimore, MD 21218, USA
[2] Sterrenkundig Observatorium, Universiteit Gent, Gent, Belgium
[3] Department of Astronomy & Astrophysics, University of California San Diego, 9500 Gilman Drive, La Jolla, San Diego, CA 92093, USA
[4] Department of Physics & Astronomy, Johns Hopkins University, 3400 N. Charles Street, Baltimore, MD 21218, USA
[5] Center for Interdisciplinary Exploration and Research in Astrophysics (CIERA) and Department of Physics and Astronomy, Northwestern University, 1800 Sherman Avenue, Evanston, IL 60201, USA
[6] Astronomy Department, University of Washington, Seattle, WA 98195, USA
[7] Center for Computational Astrophysics, Flatiron Institute, 162 Fifth Avenue, New York, NY 10010, USA
[8] Department of Astronomy, University of California, 501 Campbell Hall #3411, Berkeley, CA 94720-3411, USA
[9] Raytheon; 1151 E. Hermans Road, Tucson, AZ 85756, USA
[10] Steward Observatory, University of Arizona, 933 N. Cherry Avenue, Tucson, AZ 85719, USA
*Received 2024 April 3; revised 2024 October 21; accepted 2024 November 2; published 2025 January 7*

## Abstract

We present a catalog of individual stellar and dust extinction properties along close to 500,000 sight lines in the southwest bar of the Small Magellanic Cloud (SMC). The catalog is based on multiband Hubble Space Telescope photometric data spanning near-ultraviolet to near-infrared wavelengths from the Small Magellanic Cloud Investigation of Dust and Gas Evolution survey (SMIDGE) covering a 100 × 200 pc area. We use the probabilistic technique of the Bayesian Extinction And Stellar Tool (BEAST) to model the spectral energy distributions of individual stars in SMIDGE and include the effects of observational uncertainties in the data. We compare BEAST-derived dust extinction properties with tracers of the interstellar medium, such as the emission from the $^{12}$CO (2–1) transition ($I$(CO)), the dust mass surface density ($\Sigma_{\rm dust}$) from far-IR emission, the H I column density ($N$(H I)) from the 21 cm transition, and the mass fraction of polycyclic aromatic hydrocarbons (PAHs; $q_{\rm PAH}$, derived from IR emission). We find that the dust extinction ($A(V)$) in the SMIDGE field is strongly correlated with $\Sigma_{\rm dust}$ and $I$(CO), and less so with $N$(H I) and $q_{\rm PAH}$, and suggest potential explanations. Our extinction measurements are also sensitive to the presence of the 2175 Å bump in the extinction curve toward UV bright stars. While most do not show evidence for the bump, we identify ∼200 lines of sight that are 2175 Å bump candidates. Furthermore, we find distinct structures in the dust extinction–distance distributions that provide insights into the 3D geometry of the SMC.

*Unified Astronomy Thesaurus concepts:* Interstellar dust (836); Interstellar dust extinction (837); Interstellar medium (847); Small Magellanic Cloud (1468); Hertzsprung Russell diagram (725); Magellanic Clouds (990); Dwarf galaxies (416); Galaxy structure (622); Spectral energy distribution (2129); Stellar properties (1624)

## 1. Introduction

The study of interstellar dust and its interaction with its environment is crucial for our understanding of dust formation and evolution, interstellar medium (ISM) physics, and galaxy evolution. The Small Magellanic Cloud (SMC) is especially well suited for studying dust at low metallicity thanks to our ability to spatially resolve individual stars, which can serve as background sources for measuring extinction. With a present-day metallicity of $1/5\,Z_\odot$ (R. J. Dufour 1984; S. C. Russell & M. A. Dopita 1992; W. R. J. Rolleston et al. 1999, 2003), the SMC is also important as one of the few places where low-metallicity extinction can be studied at parsec-scale resolution.

The amount and type of dust in a galaxy is strongly dependent on metallicity. For example, dust and gas properties appear to change dramatically at a metallicity somewhat above the SMC's in terms of dust-to-gas ratios (e.g., U. J. Sofia et al. 2006;

S. Zhukovska et al. 2008; A. K. Leroy et al. 2011; R. S. Asano et al. 2013; A. Rémy-Ruyer et al. 2014; J. Roman-Duval et al. 2014; R. Feldmann 2015; P. De Vis et al. 2019), the dust mass fraction in the form of polycyclic aromatic hydrocarbons (PAHs; e.g., C. W. Engelbracht et al. 2005; B. T. Draine et al. 2007; K. M. Sandstrom et al. 2010; J. Chastenet et al. 2019; G. Aniano et al. 2020), the proportions of silicate versus carbon dust grains (F. Galliano et al. 2011; J. Chastenet et al. 2017; J. Roman-Duval et al. 2022), the UV extinction (e.g., G. C. Clayton et al. 2003; K. D. Gordon et al. 2003, hereafter G03), and the CO-to-H$_2$ conversion factor (e.g., A. D. Bolatto et al. 2013). Extinction curves—representing the wavelength dependence of dust extinction—in low-metallicity environments are notably different from those in the Milky Way (MW). For example, the majority of the observed SMC extinction curves show a steep UV rise and a weak or absent 2175 Å bump (J. Lequeux et al. 1982; K. D. Gordon & G. C. Clayton 1998; G03; K. D. Gordon et al. 2024; J. Maíz Apellániz & M. Rubio 2012), in contrast to the MW UV extinction curve, which shows a bump and a relatively shallow far-ultraviolet (FUV) rise (E. L. Fitzpatrick 1999). At the same time, in the Large Magellanic Cloud (LMC) (∼1/2 $Z_\odot$), the LMC2 supershell sample near the 30 Doradus star formation region shows an intermediate behavior (K. A. Misselt et al. 1999; G03). While it is likely that these differences hold clues about how the dust in such environments behaves and evolves, metallicity is not the only factor responsible for dust extinction properties. As an example of this complexity, not every SMC sight line lacks the 2175 Å bump (G03; MR12), nor does every MW sight line have it (L. A. Valencic et al. 2003, 2004; D. C. B. Whittet et al. 2004). The LMC likewise shows internal variation between the LMC2 supershell and the rest of the LMC (G03).

Understanding dust properties at low metallicities also affects our ability to make reliable conclusions about star formation histories, especially in terms of the cosmic star formation rate (SFR). For example, calculating the SFR of distant galaxies often depends on measurements of the UV-continuum emission from an assumed stellar initial mass function (IMF) and dust content (S. J. Lilly et al. 1996; P. Madau & M. Dickinson 2014). Since the peak of star formation in the Universe occurred at a metallicity much lower than the present-day MW's (S. Genel et al. 2014; P. Madau & M. Dickinson 2014; M. Song et al. 2016; Y. Harikane et al. 2022; C. T. Donnan et al. 2023), it is essential to characterize low-metallicity dust properties and content.

Observational and theoretical studies of the ISM show that dust extinction is an extremely important diagnostic of the ISM (C. F. McKee & E. C. Ostriker 2007; A. S. Hill et al. 2008; J. Kainulainen et al. 2009; P. Hennebelle & E. Falgarone 2012; J. J. Dalcanton et al. 2015, etc.). For example, extinction mapping can be used effectively to probe galactic processes such as the collapse of molecular clouds to form stars or to find the relationship between emission from molecular gas and the dust column density. As a tracer that does not depend on uncertainties in the far-IR opacity, dust extinction analysis based on photometric observations can additionally address whether the dust-to-gas ratio varies with phase (J. Roman-Duval et al. 2019, 2022; C. J. R. Clark et al. 2021, 2023). Dust extinction maps of $A(V)$ can also be a powerful tool to understand how the carbon monoxide (CO) emission and $A(V)$ correlate at low metallicity, and in turn understand how much shielding CO needs to form and where there is CO-dark $H_2$ (e.g., J. L. Pineda et al. 2013; W. D. Langer et al. 2014; J. Roman-Duval et al. 2014).

Interpreting variations in dust and stellar properties on parsec scales where individual molecular clouds can be studied and identified (by e.g., Atacama Large Millimeter/submillimeter Array, etc.) is an increasingly powerful tool in astronomy. For example, distinct dust extinction curve properties have been observed for sight lines within the same molecular cloud in the SMC (J. Maíz Apellániz & M. Rubio 2012, hereafter MR12), implying different interstellar dust composition and dust distribution within a few parsecs. Knowing what causes these variations is fundamental to understanding the nature of dust and, subsequently, to properly interpret astronomical observations.

Photometric data of resolved stellar populations can serve to compare dust extinction properties to those obtained via modeling of the IR dust emission using dust grain models (B. T. Draine et al. 2014; J. J. Dalcanton et al. 2015; Planck Collaboration et al. 2016). Such analyses have shown that emission-based dust mass surface density measurements may overestimate the dust mass surface density by a factor of ~2 compared to extinction measurements. This effect is due to an underestimate of the dust grain emissivity based on models calibrated to observations of MW midplane sight lines (e.g., see Section 5 in Planck Collaboration et al. 2016). Photometry-based measurements of dust extinction can therefore be used to independently calibrate dust grain models.

Ground-based surveys have shown the promise of dust extinction mapping. However, they have been limited in their sensitivity and also in their power to resolve faint stars, especially in crowded regions. The Hubble Space Telescope (HST) can be used to overcome these limitations. There have only been a handful of surveys providing deep color–magnitude diagrams (CMDs; down to F110W Vega magnitude of ~26) in the SMC, and they have been for very small fields in a limited number of HST photometric bands (M. Cignoni et al. 2013; D. R. Weisz et al. 2013). Additionally, most of the current surveys either operate in a limited wavelength range, are conducted on large angular scales limiting the number and type of stars that can be detected, or lack the observational depth of the HST.

A number of surveys present catalogs of the SMC's stellar or dust properties, or both. D. Zaritsky et al. (1997, 2002) provided a stellar catalog and a large-scale extinction map in their Magellanic Clouds Photometric Survey using $U$, $B$, $V$, and $I$ stellar photometry observed with the Great Circle Camera on the Las Campanas Swope telescope. P. Massey & A. S. Duffy (2001) cataloged O and B stars in the SMC. C. Alcock et al. (1997) and A. Udalski et al. (1998) focused on optical microlensing surveys (Optical Gravitational Lensing Experiment and macHO). The VISTA Magellanic Survey survey (M. R. L. Cioni et al. 2011) used the VISTA telescope to obtain near-infrared $YJK_S$ wide-field photometry to find distances and 3D geometry from Cepheids and red clump (RC) stars, and calculated reddening maps and star formation histories (S. Rubele et al. 2018). The SMASH survey (D. L. Nidever et al. 2017) mapped both Magellanic Clouds and focused on low surface brightness stellar populations to derive star formation histories and map extended stellar structures. The Gaia Collaboration (Gaia Collaboration et al. 2018, 2021) surveyed the full sky in the direction of the SMC and produced catalogs of stellar and dust extinction properties. These existing catalogs, however, result only in low-resolution maps of dust extinction that do not reveal dust extinction properties on subparsec scales, or have large uncertainty on stellar parameters.

To understand the relationships among stellar, dust, and gas properties on the scales of individual star-forming and/or molecular cloud regions, we bring together the rich information encoded in deep, high spatial resolution, near-UV to near-IR HST photometry of half a million stars in the SMC southwest (SW) bar, in the main body of the galaxy. Our goal is to obtain a catalog of individual stellar and dust properties from fitting the spectral energy distributions (SEDs) of a great variety of stellar types ranging from blue supergiants to red dwarfs. In addition, we compare the inferred stellar and dust extinction properties to tracers of the ISM to identify any correlations in search of the drivers of dust properties in a low-metallicity ISM, how dust and stellar properties may be related to the SMC 3D galactic geometry, and propose additional science questions that may be explored from our results.

Our analysis relies on the open-source Bayesian Extinction And Stellar Tool (BEAST; K. D. Gordon et al. 2016, hereafter G16).[11] The BEAST is a probabilistic approach to fit the observed photometric stellar SED of sources over a range of wavelengths while accounting for the effects of observational uncertainties and bias in all signal-to-noise ratio (S/N) measurements. Since the focus of this work is on the SMC dust properties and their potential correlation with the ISM, we rely on the BEAST's strength to incorporate a realistic interstellar extinction to address stellar and dust properties in tandem.

---

[11] BEAST Fitting, v2.0.1: https://github.com/BEAST-Fitting/beast.

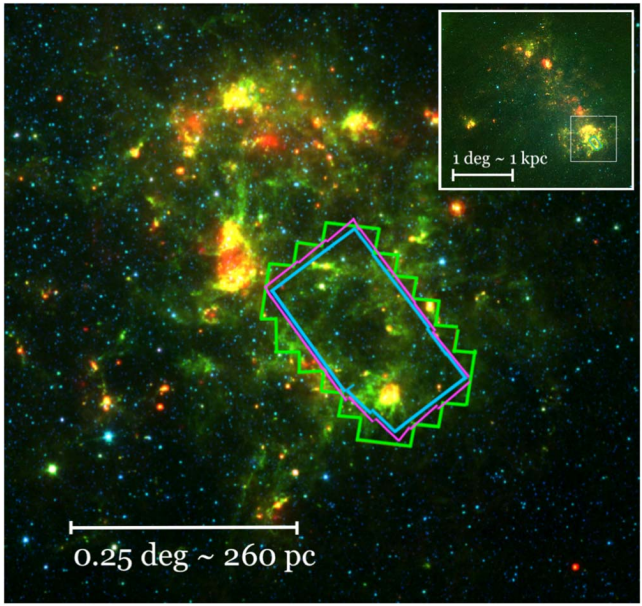

**Figure 1.** The SMIDGE survey field (YMJ17, YMJ21) covers a 12′ × 6′.5 region in the SW bar of the SMC, which is shown here as a Spitzer Space Telescope composite image in 3.6, 8, and 24 μm from the SAGE SMC survey (K. D. Gordon et al. 2011). The inset shows the galaxy's bar/body (center right) and wing region (left). The imaging footprint of the HST's cameras are shown in green (ACS/WFC), magenta (WFC3/UVIS), and blue (WFC3/IR). In red for reference is the location of the N13 SMC nebula.

In Section 2, we describe the Small Magellanic Cloud Investigation of Dust and Gas Evolution survey (SMIDGE) HST observations and ancillary data used. Our SED fitting technique is described in Section 3. Section 4 describes the stellar and dust properties derived for the SMIDGE field. A sample of the catalog containing these properties can be found in Table 4. In Section 5, we discuss our results, and finally we conclude with Section 6, describing future work to be based on this initial analysis.

## 2. Data

We use data from the SMIDGE survey (GO-13659), which covers a 12′ × 6′.5 (∼200 pc × 100 pc) contiguous region in the SW bar of the SMC, as shown in Figure 1. SMIDGE consists of deep, nine-band HST WFC3/UV imaging in F225W, F275W, and F336W, ACS imaging in F475W, F550M, F658N, and F814W, and WFC3/IR imaging in the F110W and F160W photometric bands. The photometric catalog includes ∼700,000 stars that have been subject to the quality cuts described in P. Yanchulova Merica-Jones et al. (2017, 2021, hereafter YMJ17 and YMJ21).

We impose quality cuts in the S/N, sharpness, roundness and crowding parameters from the DOLPHOT photometry processing pipeline (A. E. Dolphin 2002; A. Dolphin 2016) to eliminate low-quality measurements such as diffraction spikes, spurious detections, and low S/N sources. The following thresholds are implemented for the Advanced Camera for Surveys/Wide Field Camera (ACS/WFC) filters: sharpness squared ⩽0.1, roundness ⩽1.5, and crowding ⩽1.1. This process aims to remove poorly detected sources or sources that are not actual stars, which would not be well fit with our method. We do not remove foreground stars, but our quality cuts do remove most diffraction spikes. We additionally manually mask diffraction spikes that may remain after the quality cuts. In Section 3.1, we go into detail about how we fit the stellar SED of each source using stellar evolutionary and atmosphere models. This process, together with modeling the distance to each star, will cause stars that may be a part of the foreground to have a poor fit of their photometric SED flux and then be excluded by our postprocessing quality cuts on $\chi^2$. Section 3.4 describes how we employ a noise model to simulate observational effects and additionally assess the quality of the photometry. For complete details of the SMIDGE imaging footprint, HST camera and filter selections, photometric processing, and data culling, we refer the reader to YMJ17.

In this work we use observations in nine photometric bands—WFC3/UVIS imaging in F225W, F275W, and F336W, ACS imaging in F475W, F550M, F658N, and F814W, and WFC3/IR imaging in F110W and F160W. We trim the original catalog used in YMJ17 and YMJ21 to only include stars with a nonzero flux in all seven wide HST bands. This selection does not require a significant detection in each band, but requires sources to lie within the observed footprint for all passbands, imposing the condition that only stars that lie within the smallest footprint—that of the WFC/IR camera—are included (see Figure 1). This brings the final number to 464,765 individual stars, compared to 698,372 stars in the original YMJ17 and YMJ21 catalog. To illustrate the depth and wavelength coverage of the SMIDGE data, we show CMDs of the observations in six bands in Figure 2.

### 2.1. Ancillary Data

To study correlations between the SMIDGE stellar and dust properties and gas tracers, we make use of ancillary data consisting of the following observations: the APEX telescope $^{12}$CO (2–1) emission map at 7″ resolution (H. P. Saldaño et al. 2023); the neutral hydrogen (H I) emission combined data from the Australian Square Kilometer Array Pathfinder and Parkes Telescope (N. M. Pingel et al. 2022) at a Nyquist sampled resolution of ∼30″; maps of the mass fraction of PAHs, and the dust mass surface density ($\Sigma_{dust}$) at ∼36″ from J. Chastenet et al. (2019) based on the Spitzer Legacy SAGE SMC survey (K. D. Gordon et al. 2011) and the Herschel HERITAGE Key Project (M. Meixner et al. 2013) in the 3.6–70 μm (mid-IR) and the 100–500 μm (far-IR) wavelength range, respectively.

## 3. Model

### 3.1. Fitting Method Overview

We fit the SMIDGE observations with the BEAST (G16), which is built upon several components. Fundamentally, the BEAST infers the intrinsic properties of individual stars and of the intervening dust toward single sight lines. The underlying technique relies on a probabilistic modeling of the photometric SED of stars at a range of wavelengths, where the Bayesian framework provides a specific method by which separate models can be combined and evaluated using a grid-based approach. This process is described in detail in G16, which used the BEAST to fit observations from the Panchromatic Hubble Andromeda Treasury (J. J. Dalcanton et al. 2012). Briefly, the BEAST uses a multivariate Gaussian distribution to evaluate the probability of the model parameters given a set of photometric measurements for a single star and any prior knowledge of the model parameters from additional information.

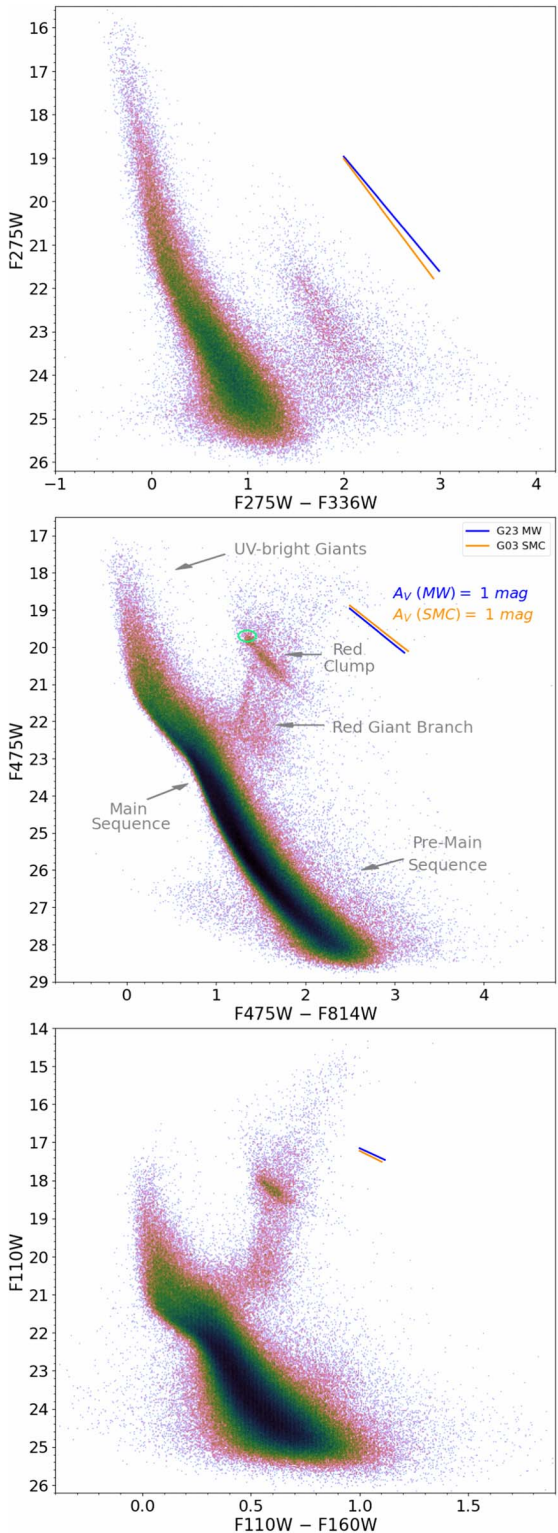

**Figure 2.** The SMIDGE CMDs based on the Basic Cuts catalog show the rich multiband information where distinct stellar populations and the effects of dust are visible across the UV (top), optical (middle), and IR (bottom) CMDs. The elongation of the RC illustrates the combined effects of dust and the SMC distance spread on stellar color (reddening) and magnitude (extinction). Vectors representing 1 mag of extinction in the $V$ band for MW- and SMC-type dust extinction from the K. D. Gordon et al. (2023) and G03 models are shown in blue and orange, respectively (arrows are omitted for clarity of vector separation).

Here, we use $N = 7$ photometric measurements of each source, and our parameter grid has $D = 7$ dimensions with one dimension for each parameter: stellar age ($t$), initial stellar mass ($M_{ini}$), metallicity ($Z$), dust column ($A(V)$), the extinction in the Johnson $V$ band ($A(V)$), which traces the dust column), the average dust grain size ($R(V)$, which is the ratio of extinction, $A(V)$, to selective extinction, $E(B − V)$), $f_\mathcal{A}$, which is the mixture coefficient indicating the fraction of MW- and SMC-type dust (with and without a 2175 Å bump) (G16), and distance. The parameter grid range and resolution are listed in Table 1.

The model SEDs are computed for each set of parameters to predict the intrinsic stellar spectrum. We then incorporate the effects of dust on the stellar spectrum, and finally compute the full extinguished stellar spectrum by integrating it over the transmission curve of each relevant HST photometric band. The BEAST determines the best-fit values from the 7D posterior probability distribution functions (pPDFs). It also calculates the 50% model, or the median, along with the range of models within a 1$\sigma$ uncertainty measured from the 67% width of the 1D pPDFs.

Figure 3 illustrates how the BEAST fits the SEDs of an individual star. The SED shown comes from a source located in the quiescent SMC B1-1 molecular cloud (M. Rubio et al. 1993; J. Lequeux et al. 1994; star 11 in MR12). Our fits indicate that this is a hot ($\log(T_{eff}) = 4.3$), mildly reddened star with $A(V) \sim 0.8$ mag and $\log(g) \sim 4.17$. $\log(T_{eff})$ and $\log(g)$ indicate a B2 V spectral-type based on the mapping between these stellar parameters and the stellar spectral type (see Table 15.8 in A. N. Cox 1999 and Tables B.3 and B.4 in R. O. Gray & C. J. Corbally 2009). The fit shows a sharply peaked $\log(T_{eff})$, $\log(g)$, and $\log(L)$ probability distribution with small uncertainties for a luminous blue main-sequence (MS) star. Age and mass are similarly well constrained for this individual example. The derived metallicity ($\log(Z) = −2.25$) corresponds to $Z = 0.37 Z_\odot$. The role of $R(V)$ and $f_\mathcal{A}$ is described in detail in Section 3.3. In summary, $R(V)$ is related to the average dust grain size along a sight line (J. A. Cardelli et al. 1989), which dictates the shape of the extinction curve, while $f_\mathcal{A}$ models the fraction of MW-type ($f_\mathcal{A} \sim 1$) and SMC-type ($f_\mathcal{A} \sim 0$) dust extinction (G16). The two-parameter relationship ($R(V)$, $f_\mathcal{A}$) is based on the average extinction behavior in the MW and in the Large and Small Magellanic Clouds.

While the BEAST can be used to study individual stars with quality measurements, it is tailored for large photometric surveys of resolved stellar populations where crowding and photon noise may contribute significantly to the uncertainty in the measured flux and bias in each band. This feature allows for the statistical interpretation of properties of stellar populations and the intervening ISM where the BEAST can accurately derive the intrinsic physical properties in all S/N regimes. The model is described in detail in Section 3.4.

Additionally, the fitting process includes priors in the calculation to account for knowledge about the stellar and

Table 1
BEAST Physics Model Grid Parameters

| Description | Unit | Min | Max | Resolution | Prior |
|---|---|---|---|---|---|
| Stellar age, $\log(t/\mathrm{yr})$ | dex | 6.0 | 10.13 | 0.1 dex | Constant SFR |
| Stellar mass (initial) | $\log(M_{\mathrm{ini}}/M_\odot)$ | −1.1 | 2.3 | Variable | Kroupa IMF |
| Stellar metallicity | $\log(Z/Z_\odot)$ | −2.1 (0.8% $Z_\odot$) | −0.3 (50% $Z_\odot$) | 0.3 dex | Flat |
| Dust column, $A(V)$ | mag | 0.01 | 10 | 0.05 mag | Flat |
| Dust grain size, $R(V)$ | ... | 2.24 | 5.74 | 0.5 | Peaked at 2.74[a] |
| Dust mixture coefficient, $f_\mathcal{A}$ | ... | 0.0 | 1.0 | 0.1 | Peaked at 1[a] |
| Distance | kpc | 47 | 77 | 2.5 | Flat over $d_{\mathrm{center}} \pm 15$ kpc range[b] |

**Notes.**
[a] The prior is defined by the distribution in 2D $R(V)$ vs. $f_\mathcal{A}$ space dictated by the mixture model for dust extinction. The resulting projection in 1D space is a prior weighted toward $f_\mathcal{A} \sim 1$ and $R(V) \sim 2.74$. See Figure 7 of G16.
[b] SMC center at 62 kpc (V. Scowcroft et al. 2016).

dust properties from previous studies. The respective priors are discussed at the beginning of Sections 3.2 and 3.3, and are also listed in Table 1. For example, we use a flat prior for the SMC distance over the range of 47–77 kpc. We take 62 kpc (V. Scowcroft et al. 2016) as the central value of the distance grid, while the spread in measured SMC distances is accounted for by the distance range itself. As a final fitting step, the 7D pPDF is calculated, implementing the priors placed on the model parameters.

### 3.2. Stellar Model

Our stellar model SED grid is built upon a combination of stellar evolutionary and atmosphere models. The fundamental stellar parameters are initial stellar mass, $M_{\mathrm{ini}}$, stellar age, $t$, and metallicity, $Z$. These parameters map directly to the stellar effective temperature ($T_{\mathrm{eff}}$), surface gravity ($\log(g)$), and luminosity ($L$). We use the stellar atmosphere grids of F. Castelli & R. L. Kurucz (2003) and T. Lanz & I. Hubeny (2003, 2007); details of how the BEAST performs the merger are described in Section 3.1 of G16. For SMIDGE, we use the PARSEC stellar evolutionary tracks (from the CMD 3.7 web interface[12]) of A. Bressan et al. (2012), a P. Kroupa (2001) IMF, and a metallicity range of ∼0.008–0.5 $Z_\odot$ (J.-K. Lee et al. 2005). For details on the stellar atmosphere grid choices, see Section 3.1 of G16.

The BEAST fitting process has known limitations since, while the code is in active development, it does not currently include models for some types of stars, such as binaries and asymptotic giant branch (AGB) stars. This limitation is likely to affect the precision of the recovery of secondary dust parameters, which are more strongly influenced by changes in the SED shapes. As a result, nonphysical fits may be evident in some parts of the CMD. We discuss this possibility in more detail in Section 4.5.

### 3.3. Dust Extinction Model

One of our goals is to quantify the amount and type of dust in the SMC SW bar by measuring the distribution of extinction curves toward many sight lines in this region. We want to fit the SMIDGE data for three dust parameters: $A(V)$, $R(V)$, and $f_\mathcal{A}$.

To model $R(V)$ and $f_\mathcal{A}$, we use a dust extinction model that describes the observed variation among MW and SMC extinction curves. While most MW sight lines show an $R(V)$-dependent extinction curve (that is, extinction curves follow a family of curves that can be described by this parameter; J. A. Cardelli et al. 1989; E. L. Fitzpatrick 1999; K. D. Gordon et al. 2023), most extinction curves in the Magellanic Clouds do not follow this $R(V)$ dependence (G03; K. D. Gordon et al. 2024). Notable differences exist in the measured SMC extinction curves, particularly in the UV part of the spectrum, where in the SMC bar there is an absence of a 2175 Å bump and a steeper rise in the FUV. Generally, extinction curves in the Local Group show a wide variation of properties that fall somewhere between the MW and the SMC extinction curve properties (G03; G. C. Clayton et al. 2015). To address this variation and to select appropriate priors on $R(V)$ and $f_\mathcal{A}$, G16 proposed a mixture model for the dust extinction where the coefficient $f_\mathcal{A}$ accounts for the fraction of MW-type dust extinction and $1 - f_\mathcal{A}$ accounts for the fraction of SMC-type dust extinction:

$$\frac{A_\lambda}{A_V} = f_\mathcal{A}\left[\frac{A_\lambda}{A_V}\right]_{\mathcal{MW}} + (1 - f_\mathcal{A})\left[\frac{A_\lambda}{A_V}\right]_{\mathcal{SMC}}. \quad (1)$$

The extinction's dependence on UV to near-IR wavelengths reveals a variety of dust column-normalized extinction curves, $A(\lambda)/A(V)$ (where $A_\lambda$ is the extinction at wavelength $\lambda$). $A(\lambda)/A(V)$ is calculated using E. L. Fitzpatrick et al. (2019), which takes into account passband effects in optical and NIR extinction curve measurements. The model we use describes a stars observed monochromatic flux after passing through a column of dust, where the full model for the flux of a star at distance $d$ is

$$F_i^{\mathrm{Mod}} = \frac{L_\lambda(\theta_{\mathrm{star}}) D_{\lambda,\mathrm{dust}}(\theta_{\mathrm{dust}})}{4\pi d^2}, \quad (2)$$

where $L_\lambda(\theta_{\mathrm{star}})$ is the wavelength-dependent luminosity, or the SED, of a single star with model parameters $\theta$, and

$$D_{\lambda,\mathrm{dust}}(\theta_{\mathrm{dust}}) = 10^{-0.4\,A(V)\,k_\lambda\,(R(V),f_\mathcal{A})} \quad (3)$$

is the dust mixture model with dust parameters $\theta_{\mathrm{dust}} = \{A(V), R(V), f_\mathcal{A}\}$. The $(R(V), f_\mathcal{A})$ parameter combination describes the shape of the extinction curve. After we establish this SED model, we compare it with photometric observations by computing model fluxes in the same bands as the observations. For details about the BEAST fitting procedure, see Sections 3.2 and 3.3 of G16.

---
[12] http://stev.oapd.inaf.it/cgi-bin/cmd

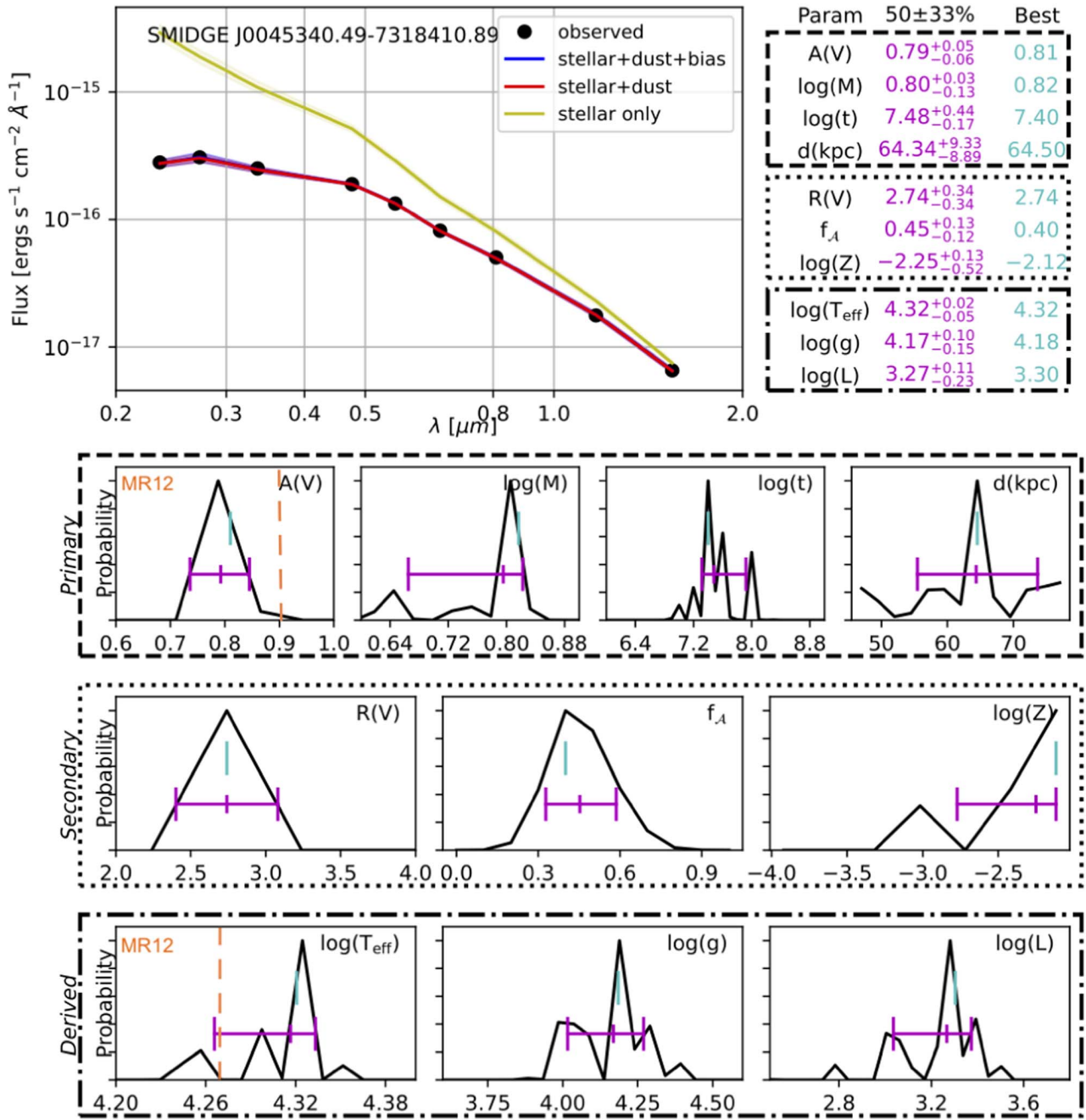

**Figure 3.** The BEAST fitting results for an individual star in SMC B1-1. The parameters in order from top left to bottom right are: $A(V)$ (mag), initial stellar mass ($M_\odot$), stellar age (Gyr), distance (kpc), $R(V)$, $f_\mathcal{A}$, metallicity ($Z/Z_\odot$), temperature (K), surface gravity (cm s$^{-2}$), and luminosity ($L_\odot$). This is a B-type blue giant star. Using HST STIS spectra and optical/NIR photometry, MR12 measure a strong 2175 Å bump for this star (star 11 in their analysis), as do K. D. Gordon et al. (2024), and the BEAST fits show $f_A \sim 0.45$, or a dust mixture with ∼50% MW-type dust (containing a 2175 Å bump). The dashed orange line shows the $\log(T_{\rm eff})$ and $A(V)$ values derived by MR12. The comparison between the temperature for the star derived by the BEAST and by MR12 can also be seen in Figure 6, along with the rest of the MR12 stars.

Figure 4 shows an example of the mixture model and how it can smoothly interpolate between SMC-type dust ($R(V) = 2.74$; $f_\mathcal{A} \sim 0$; G03) and MW-type diffuse ISM dust ($R(V) = 3.1$; $f_\mathcal{A} \sim 1$; E. L. Fitzpatrick 1999). This illustration of the mixture model results in a set of extinction curves overlaid on the SMIDGE filter band-pass functions. It can be seen that the flux in the SMIDGE F225W and F275W bands helps us detect variations in $R(V)$ and $f_\mathcal{A}$ to varying degrees, and that generally we are more sensitive to variations in $R(V)$ (dashed-line curves) than in $f_\mathcal{A}$ (solid-line curves). To be truly sensitive to the strength of the 2175 Å bump, however, in a future survey we would need observations in an FUV photometric band, such as, for example, HST's WFC3 or ACS FUV long-pass filters.

To choose physically reasonable priors in our fitting, we implement a mixture model based on the dust extinction curves by E. L. Fitzpatrick et al. (2019) and K. D. Gordon et al. (2003, 2024). We chose a prior in the 2D $R(V)$ versus $f_\mathcal{A}$ space, where $R(V)$ spans the range $2.24 < R(V) < 5.74$, to ensure the range observed in the literature, and also that the equal spacing of grid points allows for the observed SMC average $R(V) = 2.74$ (G03). The range of $f_\mathcal{A}$ encompasses the full observed range from a purely MW-type dust ($f_\mathcal{A} \sim 1$, with a 2175 Å bump) to a purely SMC-type dust ($f_\mathcal{A} \sim 0$, no 2175 Å bump). Both $R(V)$ and $f_\mathcal{A}$ are fit with equally spaced grid points. The mixture model results in the 2D $f_\mathcal{A}$ versus $R(V)$ parameter space not being completely filled, and in a 1D projection where the priors are weighted toward $f_\mathcal{A} \sim 1$ and $R(V) \sim 2.74$ (a completely uninformed prior, that is, a prior exploring every possible combination in the 2D $f_\mathcal{A} - R(V)$ space is not a part of the model as it is nonphysical (i.e., $f_\mathcal{A} = 0$ only when $R(V) = 2.74$ by definition)). For an additional description and justification of the prior on $R(V)$ and $f_\mathcal{A}$, see Section 4.3 of G16.

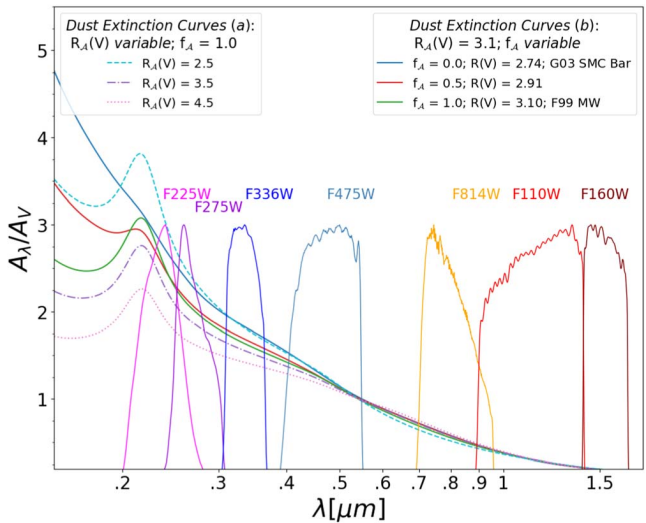

**Figure 4.** An example of the dust extinction mixture model behavior illustrates the smooth transition between dust extinction curves. Set (a) in the legend on the left shows extinction curves as a function of $R_\mathcal{A}(V)$ with $f_\mathcal{A}$ kept constant at 1.0 (purely MW-like dust extinction curve with a 2175 Å bump). Set (b) in the legend on the right shows extinction curves as a function of $f_\mathcal{A}$, where $R_\mathcal{A}(V)$ is kept constant at 3.1 (MW diffuse ISM; E. L. Fitzpatrick 1999), and $f_\mathcal{A}$ is allowed to vary. $f_\mathcal{A} = 0$ indicates a purely SMC-like dust extinction curve with $R(V) = 2.74$ (G03). The SMIDGE filter band-pass functions are overlaid and show that with the current set of filters we are more sensitive to variations in $R(V)$ than in $f_\mathcal{A}$, primarily due to the flux in F225W and F275W.

We choose to use a flat prior on $A(V)$. Though YMJ21 chose a log-normal $A(V)$ distribution based on Galactic studies of molecular clouds (A. S. Hill et al. 2008; J. Kainulainen et al. 2009), we opt for a simple, uninformative prior distribution for the present work with equally spaced $A(V)$ grid points. We will revisit this prior choice in subsequent fitting and follow-up investigations. We discuss the limitations in our fitting in Section 4.5.

### 3.4. Noise Model

We generate a noise model to simulate observational effects in the data and account for uncertainties stemming from the HST absolute flux calibration, instrument noise, and crowding in the SMIDGE field. More specifically, the purpose of the noise model is to quantify the degree to which the model SEDs built upon our grid of stellar and dust parameters differ from the SEDs of observed stars. To address the nonlinear interaction of the photon and crowding noise, the model calculates the systematic bias from the mean offset between the measured and the true flux of a source. The flux offset occurs in each filter when high stellar crowding (causing a contaminated flux measurement due to neighboring sources) may cause an under- or overestimation of the measured flux. We expect the stellar crowding effect in the SMIDGE field to be negligible due to the SMC's relative proximity and the high resolution of the HST.

We use artificial star tests (ASTs) to assess the quality of our photometry and to quantify the photon and crowding noise. ASTs can also be used to measure completeness and photometric errors due to blending. We generated ASTs using DOLPHOT by the process described in the following paragraph, in pairs of photometric bands for each HST camera, by conservatively assuming that the bands are independent of each other. This is the "Toothpick" noise model, and it allows for speed optimization when a limited number of ASTs are generated to a large observations catalog. G16 go into detail to investigate how the fit parameter uncertainties are affected by this assumption, and by the implementation of a noise model built from the full N-band ASTs to account for the covariance between photometric bands due to stellar crowding.

An input list of ~200,000 ASTs with known SEDs was generated, and ASTs were inserted into the observations using a simulated point spread function and photon noise. We then extracted the flux of these artificial stars in the simulated observations with the same photometric pipeline routine—DOLPHOT—used to generate the point source catalog of the original observations. G16 determine that stellar noise properties can vary significantly with local source density within a field. Hence, the way we determined where to place the ASTs was to calculate the number of sources within a square pixel of $5''$ with $15 \leqslant \text{F475W}$. The process was repeated numerous times in different locations of the survey area to quantify the photometric quality as a function of stellar density (number of stars per square arcsec). Measurements of the SEDs of the simulated observations allowed us to derive the uncertainty (error) and the systematic deviation (bias) for each filter. The AST analysis of the SMIDGE photometry follows closely the analysis applied to the photometry of the Panchromatic Hubble Andromeda Treasury: Triangulum Extended Region survey (B. F. Williams et al. 2021), outlined in their Section 2.3.

We generate ASTs based on the same SED grid as that of our physics model (see Table 1), at a range of ages, metallicities, and extinctions to span the color–magnitude space of the observed SEDs. The input ASTs list includes stellar fluxes with $15 \leqslant \text{F475W} \leqslant 42$ mag to ensure we go well beyond the camera's detection limit, and also due to red colors of some models for some of the older, high-metallicity stars, which may be detectable in the IR (though we conclude that in similar future analysis we can safely lower this faint limit). The noise model inherently depends on the stellar number density; therefore, we build a noise model with ASTs grouped in bins of stellar number densities using five evenly and linearly spaced bins with densities ranging from 0 to 2 stars per $\text{arcsec}^2$. We then place the artificial stars randomly within the pixels corresponding to each source density bin. The final input ASTs list contains ~200,000 sources. Using this setup, we compute the noise model for each stellar number density bin. To speed up the computation, we trim the SED grid of models unlikely to match the data.

We include absolute flux calibration uncertainties in our noise model. R. C. Bohlin et al. (2014) provide the details of this calibration to establish flux standards for HST's UV to

**Table 2**
SMIDGE BEAST Catalogs and Quality Cuts

| Catalog | Quality Cuts | # Stars | % of Total | FLAG |
|---|---|---|---|---|
| All Stars | No cuts | 464,765 | 100.0 | 0 |
| Basic Cuts | $\chi^2_{\min}$ cuts | 459,420 | 98.9 | 1 |
| High SNR | $\chi^2_{\min}$ cuts + $A(V)$ SNR $\geqslant 6$ | 44,065 | 9.5 | 2 |

**Note.** The High SNR catalog is the main catalog used in this work. The full catalog with all fit stars will be available upon publication as an HST high level science product (HLSP) at doi:10.17909/f9wx-e637 (as a part of a single .fits file). The FLAG column here and in the online catalog indicates the quality cut applied.

mid-IR bands, basing the calibration on the predicted to observed ratios of white dwarf stars from spectrophotometry. Section 4.4.1 of G16 contains further details relevant for our BEAST SED fitting.

## 4. Results

### 4.1. Quality Cuts

After we fit each individual SMIDGE star with the BEAST, we apply cuts with several quality criteria to the final catalog. The first criterion is based on the $\chi^2$ of the best-fit model, which incorporates information about the quality of the fit between the source and the physics grid combined with the noise model. For the Toothpick BEAST noise model implemented here, we compute $\chi^2$ using a normal/Gaussian function, assuming the measurements are independent between bands. As an initial rough cut, we remove sources with $\chi^2 > 100$ (~1.0% of the total number of fit sources), as we observe that this threshold removes the tail of the distribution with clearly inferior fits to the data. Fits beyond this threshold, in the tail of the $\chi^2$ distribution, are of stars that occupy unusual spaces in the H-R diagram and appear to be attributable to sources at the edge of a parameter grid, to background and/or faint sources, and to sources with observed SEDs that do not conform to expected extincted stellar SEDs.

We present two main catalogs: one is the complete catalog composed only of sources with $\chi^2_{\min} \leqslant 100$ ($N = 459,420$, or 98.9% of the total), which we call the Basic Cuts catalog; the other contains a subset of stars with highly reliable $A(V)$ fits, where the reliability is defined by a threshold in the S/N in $A(V)$ such that low $A(V)$ sources and sources with high $A(V)$ uncertainties are removed. The $A(V)$ S/N is defined as $A(V)_{p50}/\sigma$, where $\sigma = ((A(V)_{p84} - A(V)_{p16})/2$, which is the 50% model divided by the $1\sigma$ uncertainties measured from the 16–84 percentile range. We experiment with various thresholds, and determine that an S/N $\geqslant 6$ results in a mean $A(V)_{p50}$ just above 1 mag, which ensures we are examining sight lines with at least a moderate amount of dust, in excess of the MW foreground toward the SMIDGE field of $A(V) \sim 0.18$ mag (E. Muller et al. 2003; D. E. Welty et al. 2012). A catalog with this S/N in $A(V)$ consists of about 10% of the total fit sources ($N = 44,065$) and accounts only for high-quality SNR measurements of the dust column. We refer to the latter as the High SNR catalog. Both catalogs and their statistics are given in Table 2.

### 4.2. SMIDGE BEAST Catalogs

We provide the full catalog of stellar and dust extinction properties in an online version, which contains FLAG columns to indicate the quality cuts described in Table 2: (1) the full stellar catalog containing all fit sources (All Stars); (2) the Basic Cuts catalog containing all well-fit sources with $\chi^2 \leqslant 100$; and (3) the High A(V) SNR catalog containing sources with an S/N in $A(V) \geqslant 6$. In Figure 5, we present the stellar and dust extinction properties for the SMIDGE field based on these quality criteria, and also for other thresholds of the S/N in $A(V)$. The distributions are based on the 50% (median) values for each source, calculated as described in Section 3.1. We show the physics model prior for each parameter above the main histogram.

We examine results for catalogs based on various smaller and larger S/Ns in $A(V)$, where we specifically tested results for sources with an S/N in $A(V) \geqslant 3$ to evaluate an expanded sample. We also tested stricter thresholds of S/N in $A(V)$: $\geqslant 10$ and $\geqslant 15$ to evaluate any changes in the distributions. The extent of the shifts in the distributions based on samples with an increasing S/N in $A(V)$ vary from parameter to parameter. Overall, they are consistent with what we might expect when selecting dustier sight lines. Shifts are evident to higher initial stellar masses, younger ages, and higher temperatures, which are consistent with observing stars in dustier regions where hotter, younger and more massive stars tend to form and reside. By design, we see a shift in the peak of the $A(V)$ distribution to higher values, from 0.25 to 0.8 mag from the Basic Cuts to the catalog with S/N in $A(V) \geqslant 15.0$, respectively. The $R(V)$ and $f_A$ distributions also show subtle shifts, which we discuss in more detail in Section 4.4. We observe corresponding variations in the 2D distributions, which we discuss in Section 5.1.2.

To quantify the types of stars that determine the derived dust and stellar parameters, in Table 3 we provide statistics on the number of stellar population types and their corresponding stellar masses as fit by the BEAST. The stellar populations we divide the CMD into are MS, red giant branch (RGB), RC, and pre-MS (PMS) stars. The table shows the statistics for the High SNR A(V) catalog and also for the Basic Cuts catalog. The main difference between the two catalogs is that lower-mass stars with $M/M_\odot \leqslant 0.7$ contribute more significantly to the Basic Cuts catalog with 36% of all stars in this mass range, while the High SNR A(V) catalog primarily contains stars with masses in the 0.7–1.5 $M/M_\odot$ range (72%) and the $>1.5 M/M_\odot$ range (28%). This is expected as the High SNR A(V) cut tends to eliminate some lower-mass MS stars with a lower SNR in $A(V)$, which are inherently fainter and therefore may have poorer fits. Overall, MS stars contribute 89% versus 97% to the High SNR A(V) and the Basic Cuts catalogs, respectively. It is interesting to note that the combined populations of the RGB, RC, and PMS stars contribute more significantly to the High SNR A(V) catalog, comprising 7.1% of the total number of stars (and only 1.8% of the total number of stars in the Basic Cuts catalog). It is also of note that the mean derived $A(V)$ in the lower MS for the High SNR A(V) catalog is higher than the mean $A(V)$ for the same population in the Basic Cuts catalog. This trend is similarly present since the low-mass, faint stars in the lower MS have a higher $A(V)$ uncertainty.

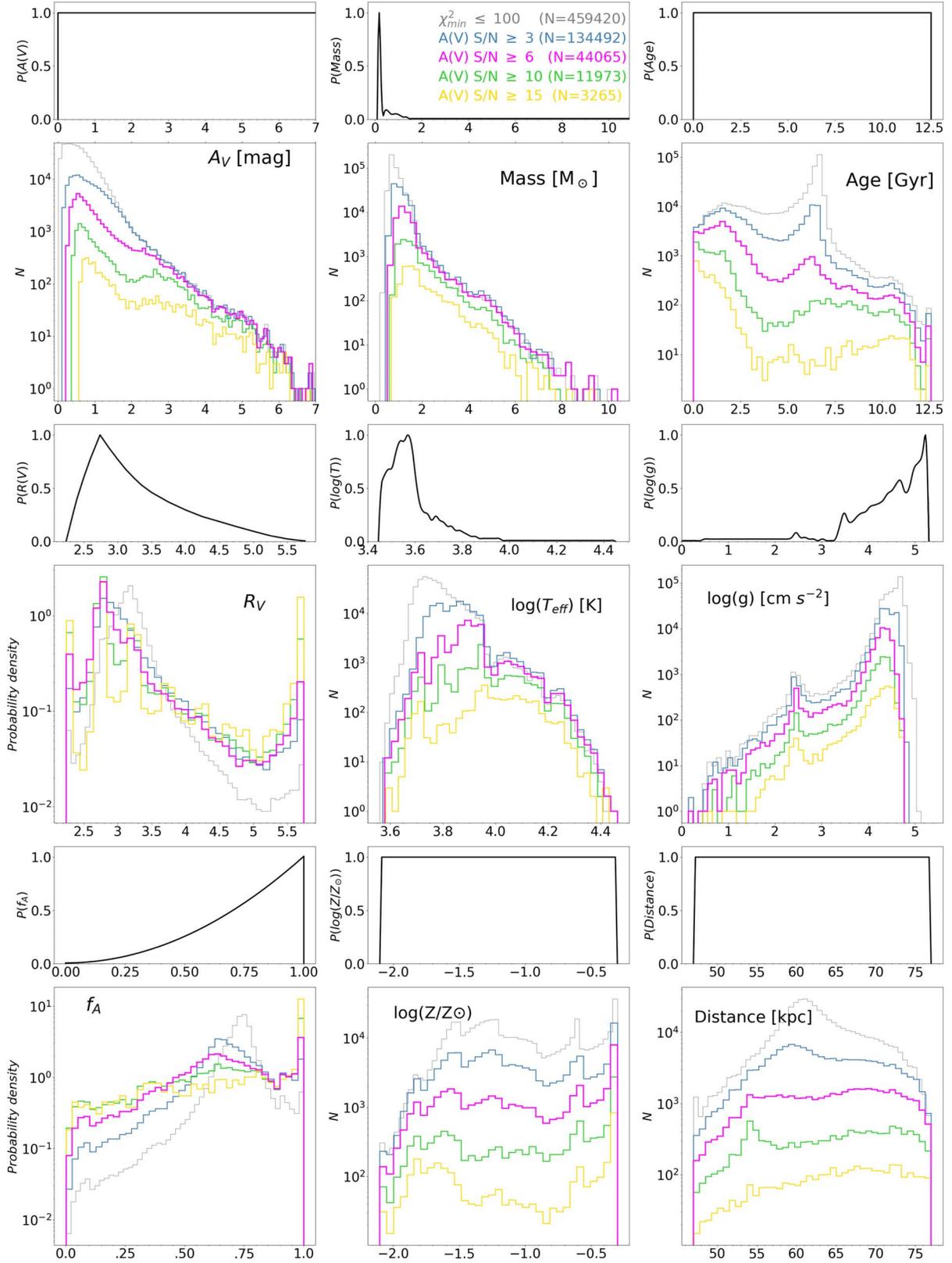

**Figure 5.** Posterior probability distribution functions (pPDF) of the SMIDGE dust and stellar parameters separated by varying S/N in $A(V)$. The full catalog (with a $\chi^2_{min}$ cut) is in gray, and the High SNR A(V) (S/N ⩾ 6) is in magenta. Several other S/N ratio thresholds are plotted in blue (S/N ⩾ 3), green (S/N ⩾ 10), and yellow (S/N ⩾ 15). The $R(V)$ and $f_A$ panels show the distribution as a probability density. The distributions are based on the median value for each fit, which is the 50% model (see Section 3.1 for details). The 1D prior for each BEAST-fit parameter based on the SMIDGE physics model grid (Table 1) is shown above each cumulative pPDF. A shift in the peak(s) of the distributions for all parameters is observed with increasing S/N threshold (discussed in Section 4.2).

**Table 3**
Color Magnitude Diagram Stellar Population Demographics

| Population | High SNR $A(V)$ Catalog ($N = 44{,}065$) | | | | Basic Cuts Catalog ($N = 459{,}420$) | | | |
|---|---|---|---|---|---|---|---|---|
| | Fraction (%) | Mass ($M/M_\odot$) | | | Fraction (%) | Mass ($M/M_\odot$) | | |
| | | $\leqslant 0.7$ | 0.7–1.5 | $>1.5$ | | $\leqslant 0.7$ | 0.7–1.5 | $>1.5$ |
| All | 100 | 69 | 31,642 (72%) | 12,354 (28%) | 100 | 164,822 (36%) | 272,881 (59%) | 21,717 (5%) |
| MS | 89 | 28 | 27,414 (62%) | 11,704 (27%) | 97 | 163,785 (36%) | 263,495 (57%) | 20,033 (4%) |
| RGB | 3.4 | 0 | 1271 (3%) | 218 (0.5%) | 0.9 | 0 | 3,763 (0.8%) | 579 (0.1%) |
| RC | 2.1 | 0 | 737 (2%) | 207 (0.5%) | 0.7 | 0 | 2,540 (0.6%) | 806 (0.2%) |
| PMS | 1.6 | 1 | 676 (2%) | 39 | 0.2 | 5 | 714 (0.2%) | 41 |

**Note.** The values represent BEAST fits and the number of stars in a CMD population within the indicated mass range. The CMD populations are MS, RGB, RC, and PMS. Percentages are given in parentheses (and are omitted when the fraction is $\ll 1$).

Table 4 shows a sample of our results for the SMIDGE stellar and dust extinction properties as derived by the BEAST. The table is a random sampling of sources in the SMIDGE field to present an example of the types of stars and fit results based on the `High SNR A(V)` catalog.

### 4.3. Stellar Properties

The effective temperature, $\log(T_{\rm eff})$, and surface gravity, $\log(g)$, derived by the BEAST allow us to infer the stellar spectral class. First, we compare the BEAST-derived physical properties to the literature values within the SMIDGE footprint. One comparison we make is with the temperature derived for stars observed via HST/Space Telescope Imaging Spectrograph (STIS) UV spectroscopy and optical/NIR photometry by MR12, who measured the extinction curve toward sight lines in the quiescent SMC B1-1 molecular cloud (M. Rubio et al. 1993; J. Lequeux et al. 1994). Figure 6 shows a good correspondence between temperatures determined from the MR12 fits to UV spectra and our BEAST fits (Figure 3 also shows the MR12 $\log(T_{\rm eff})$ and $\log(g)$ results for their star 11).

The derived stellar properties are shown in the 1D pPDF distributions in Figure 5. We start with flat priors for log(age), log(metallicity), and distance. While we adopt a constant SFR for the SMIDGE field, our fits show a stellar age distribution with several peaks. The prior on metallicity is based on a range of metallicity values for the SMC from the literature. We see several peaks in the metallicity distribution as well. The distance grid range of 47–77 kpc is well sampled, and the distribution is double peaked. In Section 5.1.2, we compare these distance results to a recent study by C. E. Murray et al. (2024a) of the SMC 3D ISM structure.

Figure 7 shows CMDs color coded by the BEAST fits for $A(V)$ and distance. While some results are intuitive based on stellar atmosphere models or the IMF, which form the basis of our fitting, we note that, for parameters for which we use flat priors, such as $A(V)$ and distance, we observe fits that have a physical explanation but also some inherent fitting biases. For example, it is intuitive that we see unreddened ($A(V) \sim 0$) stars on the blue side of CMD features, such as the MS and the RGB, and moderately or highly reddened stars toward redder colors on the CMD. At the same time, we are not certain whether the higher degree of extinction measured for stars in the lower-right region of the CMD is primarily due to PMS embedded in dust clouds (the PMS phase is somewhat challenging to identify; A. Bressan et al. 2012), or whether stars at young ages (1–5 Myr) are significantly less likely compared to older ages. We estimate that in the `High SNR A(V)` catalog the fraction of PMS stars is 1.6% (while they comprise 0.2% of the `Basic Cuts` catalog; see Table 3). The PARSEC 2.0 stellar evolutionary tracks of A. Bressan et al. (2012) that we use in our fitting include PMS stellar models, where PMS evolution specifics and uncertainties can be found in their Section 5.2.

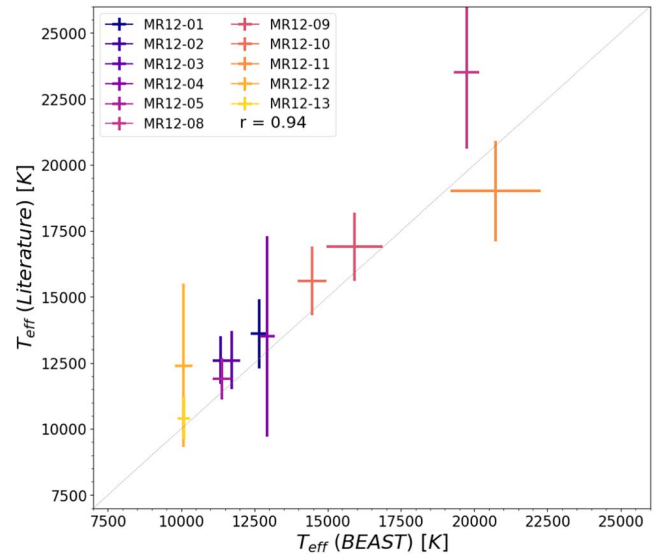

**Figure 6.** Stellar effective temperatures ($T_{\rm eff}$) derived by the BEAST compared to $T_{\rm eff}$ derived by MR12 from spectral fitting of sources within the SMC B1-1 molecular cloud complex. We see a general consistency in $T_{\rm eff}$ obtained by these two independent methods ($r = 0.94$). Additional derived parameters for the MR12 stars are presented in Table 4.

Clearly visible in the color-coded CMD in Figure 7 is the diagonal extension, or smearing, of the color and magnitude. Similar to the morphology of the RC feature (which in the absence of dust is a tight clump in CMD space, where stars reside from $\sim 1$ to 10 Gyr), this effect is caused by dust extinction and the distance spread of the SMC, as seen in the bottom panel of Figure 7 (and also in Figure 2). We also note that the distribution of the BEAST fits for the RC reproduces the combined effects of distance and reddening first discussed by YMJ17 and further explored in YMJ21. The closer (50–65 kpc) RC stars are primarily low $A(V)$, while the most reddened RC stars are more distant (65–75 kpc), which are conclusions also reached by our previous work using an

Table 4
A Sample of the SMIDGE BEAST Catalog of Stellar and Dust Properties

| Star # | R.A. (deg) | Decl. (deg) | $A(V)$ (mag) | $R(V)$ | $f_A$ | $T_{\rm eff}$ (K) | $\log(g)$ (cm s$^{-2}$) | Distance (kpc) |
|---|---|---|---|---|---|---|---|---|
| 1 | 11.3222 | −73.3680 | 0.91 ± 0.15 | 3.53 ± 0.43 | 0.77 ± 0.21 | 7129 ± 269 | 3.92 ± 0.11 | 66.9 ± 8.4 |
| 2 | 11.5294 | −73.3510 | 1.46 ± 0.24 | 3.09 ± 0.43 | 0.65 ± 0.30 | 8085 ± 547 | 4.48 ± 0.07 | 71.4 ± 4.8 |
| 3 | 11.4147 | −73.3281 | 0.74 ± 0.08 | 3.52 ± 0.51 | 0.77 ± 0.22 | 5166 ± 110 | 3.27 ± 0.12 | 62.7 ± 11.1 |
| 4 | 11.5469 | −73.2453 | 0.64 ± 0.08 | 2.72 ± 0.36 | 0.29 ± 0.35 | 8756 ± 346 | 4.32 ± 0.08 | 74.7 ± 4.2 |
| 5 | 11.3828 | −73.3254 | 0.65 ± 0.05 | 3.18 ± 0.36 | 0.79 ± 0.16 | 17517 ± 449 | 4.26 ± 0.11 | 56.6 ± 8.9 |
| 6 | 11.4795 | −73.3529 | 0.50 ± 0.05 | 5.37 ± 0.47 | 0.99 ± 0.04 | 4781 ± 102 | 2.26 ± 0.03 | 49.8 ± 1.9 |
| 7 | 11.7761 | −73.3007 | 0.71 ± 0.08 | 2.25 ± 0.18 | 1.00 ± 0.04 | 9115 ± 294 | 4.40 ± 0.05 | 74.5 ± 3.2 |
| 8 | 11.2936 | −73.3913 | 0.75 ± 0.10 | 2.67 ± 0.36 | 0.90 ± 0.18 | 8801 ± 286 | 4.25 ± 0.05 | 74.4 ± 4.0 |
| 9 | 11.3942 | −73.3055 | 2.73 ± 0.34 | 2.84 ± 0.41 | 0.59 ± 0.37 | 6067 ± 461 | 4.55 ± 0.10 | 59.0 ± 10.8 |
| 10 | 11.4896 | −73.2825 | 0.66 ± 0.06 | 5.07 ± 0.55 | 0.98 ± 0.05 | 9735 ± 288 | 4.32 ± 0.06 | 60.1 ± 6.3 |
| 11 | 11.3651 | −73.3644 | 2.29 ± 0.38 | 3.06 ± 0.40 | 0.64 ± 0.29 | 7094 ± 734 | 4.28 ± 0.14 | 55.3 ± 9.1 |
| 12 | 11.2743 | −73.4072 | 1.20 ± 0.12 | 2.74 ± 0.35 | 0.68 ± 0.34 | 7907 ± 315 | 4.36 ± 0.14 | 59.3 ± 9.9 |
| 13 | 11.5512 | −73.2238 | 1.21 ± 0.16 | 2.71 ± 0.36 | 0.35 ± 0.39 | 8598 ± 578 | 4.49 ± 0.08 | 72.6 ± 4.5 |
| 14 | 11.5614 | −73.3333 | 0.66 ± 0.06 | 2.94 ± 0.42 | 0.59 ± 0.33 | 7577 ± 137 | 4.11 ± 0.07 | 64.8 ± 7.0 |
| 15 | 11.5234 | −73.2874 | 0.36 ± 0.04 | 2.76 ± 0.35 | 0.54 ± 0.22 | 14596 ± 308 | 4.40 ± 0.12 | 56.5 ± 8.9 |
| 16 | 11.6500 | −73.2909 | 0.43 ± 0.04 | 2.74 ± 0.34 | 0.75 ± 0.19 | 12906 ± 292 | 4.45 ± 0.06 | 51.2 ± 3.3 |
| 17 | 11.2803 | −73.3437 | 0.99 ± 0.13 | 2.80 ± 0.38 | 0.63 ± 0.36 | 7476 ± 286 | 4.41 ± 0.10 | 63.4 ± 7.1 |
| 18 | 11.3065 | −73.3040 | 1.33 ± 0.18 | 3.29 ± 0.39 | 0.64 ± 0.27 | 8283 ± 459 | 4.42 ± 0.10 | 68.4 ± 6.3 |
| 19 | 11.6301 | −73.3238 | 0.44 ± 0.04 | 2.75 ± 0.35 | 0.68 ± 0.33 | 4980 ± 87 | 2.87 ± 0.07 | 48.7 ± 6.9 |
| 20 | 11.6093 | −73.2543 | 0.81 ± 0.10 | 2.75 ± 0.34 | 0.40 ± 0.33 | 8543 ± 336 | 4.16 ± 0.09 | 68.9 ± 6.9 |
| 21 | 11.5352 | −73.3624 | 0.75 ± 0.06 | 3.22 ± 0.35 | 0.38 ± 0.11 | 10909 ± 295 | 3.74 ± 0.12 | 64.4 ± 8.1 |
| 22 | 11.5655 | −73.2585 | 0.87 ± 0.13 | 2.78 ± 0.38 | 0.54 ± 0.38 | 7564 ± 340 | 4.31 ± 0.12 | 66.2 ± 9.5 |
| 23 | 11.4946 | −73.2808 | 2.43 ± 0.37 | 5.32 ± 0.62 | 0.98 ± 0.05 | 6144 ± 468 | 4.54 ± 0.10 | 58.3 ± 10.6 |
| 24 | 11.5424 | −73.2606 | 1.03 ± 0.16 | 2.81 ± 0.37 | 0.53 ± 0.36 | 8535 ± 603 | 4.41 ± 0.10 | 69.2 ± 6.4 |
| 25 | 11.2674 | −73.3855 | 0.67 ± 0.04 | 3.23 ± 0.35 | 0.78 ± 0.16 | 12966 ± 258 | 4.50 ± 0.13 | 48.0 ± 9.3 |
| 26 | 11.2962 | −73.3319 | 3.19 ± 0.29 | 5.57 ± 0.32 | 1.00 ± 0.04 | 6466 ± 421 | 4.42 ± 0.12 | 55.6 ± 9.5 |
| 27 | 11.7633 | −73.3107 | 0.40 ± 0.05 | 2.54 ± 0.37 | 0.95 ± 0.11 | 12373 ± 323 | 4.46 ± 0.09 | 70.7 ± 9.7 |
| 28 | 11.4363 | −73.3781 | 0.37 ± 0.06 | 2.74 ± 0.38 | 0.67 ± 0.36 | 8015 ± 160 | 4.33 ± 0.03 | 76.2 ± 1.5 |
| 29 | 11.5582 | −73.3598 | 1.07 ± 0.17 | 5.39 ± 0.50 | 0.98 ± 0.04 | 7041 ± 306 | 4.33 ± 0.13 | 62.7 ± 8.6 |
| 30 | 11.5300 | −73.3680 | 0.52 ± 0.04 | 2.74 ± 0.34 | 0.03 ± 0.06 | 18469 ± 411 | 4.35 ± 0.12 | 56.8 ± 9.1 |
| 31 | 11.6762 | −73.3393 | 0.58 ± 0.05 | 2.92 ± 0.41 | 0.52 ± 0.22 | 13639 ± 414 | 4.43 ± 0.08 | 64.6 ± 10.3 |
| 32 | 11.5810 | −73.2160 | 0.40 ± 0.06 | 2.61 ± 0.38 | 0.73 ± 0.27 | 12773 ± 341 | 4.36 ± 0.08 | 71.1 ± 10.8 |
| 33 | 11.5431 | −73.3706 | 1.34 ± 0.14 | 2.74 ± 0.34 | 0.54 ± 0.36 | 8737 ± 506 | 4.38 ± 0.11 | 66.8 ± 7.2 |
| 34 | 11.3786 | −73.3128 | 1.99 ± 0.18 | 3.18 ± 0.36 | 0.59 ± 0.27 | 8032 ± 431 | 4.23 ± 0.12 | 54.2 ± 9.1 |
| 35 | 11.3038 | −73.2932 | 2.60 ± 0.40 | 3.00 ± 0.52 | 0.61 ± 0.34 | 5284 ± 335 | 4.73 ± 0.06 | 66.9 ± 10.0 |
| 36 | 11.3914 | −73.3580 | 3.11 ± 0.26 | 2.70 ± 0.36 | 0.58 ± 0.39 | 6450 ± 444 | 4.34 ± 0.13 | 55.2 ± 9.3 |
| 37 | 11.3718 | −73.3088 | 1.48 ± 0.15 | 2.82 ± 0.37 | 0.65 ± 0.33 | 8648 ± 528 | 4.44 ± 0.09 | 67.6 ± 6.4 |
| 38 | 11.4115 | −73.4001 | 2.85 ± 0.29 | 2.60 ± 0.39 | 0.74 ± 0.39 | 6283 ± 450 | 4.48 ± 0.11 | 58.8 ± 10.5 |
| 39 | 11.3598 | −73.3194 | 0.38 ± 0.04 | 2.66 ± 0.37 | 0.89 ± 0.21 | 5066 ± 121 | 2.42 ± 0.04 | 65.5 ± 3.8 |
| 40 | 11.5761 | −73.2516 | 0.71 ± 0.08 | 2.85 ± 0.37 | 0.38 ± 0.31 | 8890 ± 320 | 4.24 ± 0.08 | 73.5 ± 4.7 |
| 41 | 11.4564 | −73.3137 | 2.20 ± 0.26 | 4.59 ± 0.52 | 0.94 ± 0.09 | 8945 ± 868 | 4.52 ± 0.07 | 73.1 ± 4.3 |
| 42 | 11.5577 | −73.2554 | 0.44 ± 0.05 | 4.09 ± 0.48 | 0.78 ± 0.16 | 7110 ± 124 | 4.10 ± 0.08 | 69.3 ± 6.7 |
| 43 | 11.5733 | −73.2732 | 0.42 ± 0.05 | 2.74 ± 0.35 | 0.16 ± 0.22 | 8624 ± 204 | 4.36 ± 0.02 | 74.1 ± 2.2 |
| 44 | 11.6795 | −73.2592 | 1.47 ± 0.19 | 2.77 ± 0.35 | 0.53 ± 0.37 | 7694 ± 495 | 4.34 ± 0.12 | 64.4 ± 8.8 |
| 45 | 11.4114 | −73.3718 | 0.93 ± 0.15 | 2.68 ± 0.36 | 0.76 ± 0.31 | 8680 ± 564 | 4.35 ± 0.10 | 69.0 ± 8.1 |
| 46 | 11.6068 | −73.2194 | 0.44 ± 0.06 | 2.67 ± 0.39 | 0.38 ± 0.44 | 7686 ± 187 | 4.16 ± 0.08 | 75.5 ± 6.8 |
| 47 | 11.4520 | −73.3042 | 0.40 ± 0.05 | 4.74 ± 0.35 | 0.95 ± 0.07 | 14523 ± 315 | 3.83 ± 0.04 | 54.5 ± 1.8 |
| 48 | 11.5686 | −73.3175 | 0.43 ± 0.07 | 2.69 ± 0.39 | 0.81 ± 0.30 | 7560 ± 203 | 4.21 ± 0.13 | 61.7 ± 8.1 |
| 49 | 11.3165 | −73.3734 | 2.43 ± 0.27 | 5.32 ± 0.47 | 0.99 ± 0.04 | 8348 ± 760 | 4.36 ± 0.13 | 62.5 ± 9.3 |
| 50 | 11.5737 | −73.3609 | 3.06 ± 0.25 | 3.19 ± 0.39 | 0.73 ± 0.26 | 6314 ± 414 | 4.17 ± 0.16 | 54.9 ± 9.9 |
| MR12-01 | 11.3994 | −73.3073 | 0.15 ± 0.05 | 3.22 ± 0.44 | 0.59 ± 0.32 | 12667 ± 294 | 4.15 ± 0.13 | 56.3 ± 8.7 |
| MR12-02 | 11.4055 | −73.3076 | 0.17 ± 0.05 | 3.47 ± 0.49 | 0.76 ± 0.27 | 11349 ± 279 | 4.18 ± 0.12 | 57.9 ± 9.1 |
| MR12-03 | 11.4070 | −73.3078 | 0.19 ± 0.06 | 3.19 ± 0.39 | 0.44 ± 0.28 | 11718 ± 297 | 4.16 ± 0.15 | 61.2 ± 9.7 |
| MR12-04 | 11.3906 | −73.3077 | 0.81 ± 0.04 | 2.74 ± 0.34 | 0.43 ± 0.16 | 12934 ± 256 | 4.40 ± 0.07 | 50.4 ± 4.7 |
| MR12-05 | 11.3999 | −73.3083 | 0.34 ± 0.06 | 4.28 ± 0.36 | 0.77 ± 0.13 | 11389 ± 306 | 3.95 ± 0.05 | 49.6 ± 1.8 |
| MR12-08 | 11.3889 | −73.3091 | 1.23 ± 0.04 | 3.23 ± 0.34 | 0.99 ± 0.04 | 19737 ± 435 | 4.21 ± 0.13 | 54.7 ± 9.1 |
| MR12-09 | 11.3963 | −73.3100 | 0.40 ± 0.08 | 3.34 ± 0.53 | 0.47 ± 0.32 | 15917 ± 962 | 3.90 ± 0.15 | 68.3 ± 9.6 |
| MR12-10 | 11.4046 | −73.3111 | 0.74 ± 0.05 | 3.18 ± 0.36 | 0.31 ± 0.10 | 14472 ± 489 | 3.91 ± 0.09 | 68.1 ± 8.5 |
| MR12-11 | 11.3937 | −73.3116 | 0.79 ± 0.05 | 2.74 ± 0.34 | 0.45 ± 0.13 | 20730 ± 1545 | 4.17 ± 0.13 | 64.3 ± 9.1 |
| MR12-12 | 11.3873 | −73.3121 | 0.08 ± 0.06 | 3.04 ± 0.55 | 0.51 ± 0.37 | 10085 ± 302 | 4.33 ± 0.09 | 49.8 ± 4.8 |
| MR12-13 | 11.3838 | −73.3123 | 0.09 ± 0.06 | 2.92 ± 0.43 | 0.71 ± 0.29 | 10092 ± 214 | 3.83 ± 0.04 | 65.1 ± 4.0 |

**Note.** This is a sample based on the BEAST-derived SNR$_{A_V}$ ⩾ 6 catalog consisting of 44,065 sources and also the 11 MR12 stars within the SMIDGE survey footprint, which are listed for reference (stars MR12-08, 09, 10, and 11 are the four stars with fit extinction curves by MR12, with MR12-09, 10, and 11 also fit by K. D. Gordon et al. 2024). The full catalog, as well as all other catalogs described in Table 2, will be available as an HLSP at doi:10.17909/f9wx-e637 upon publication.

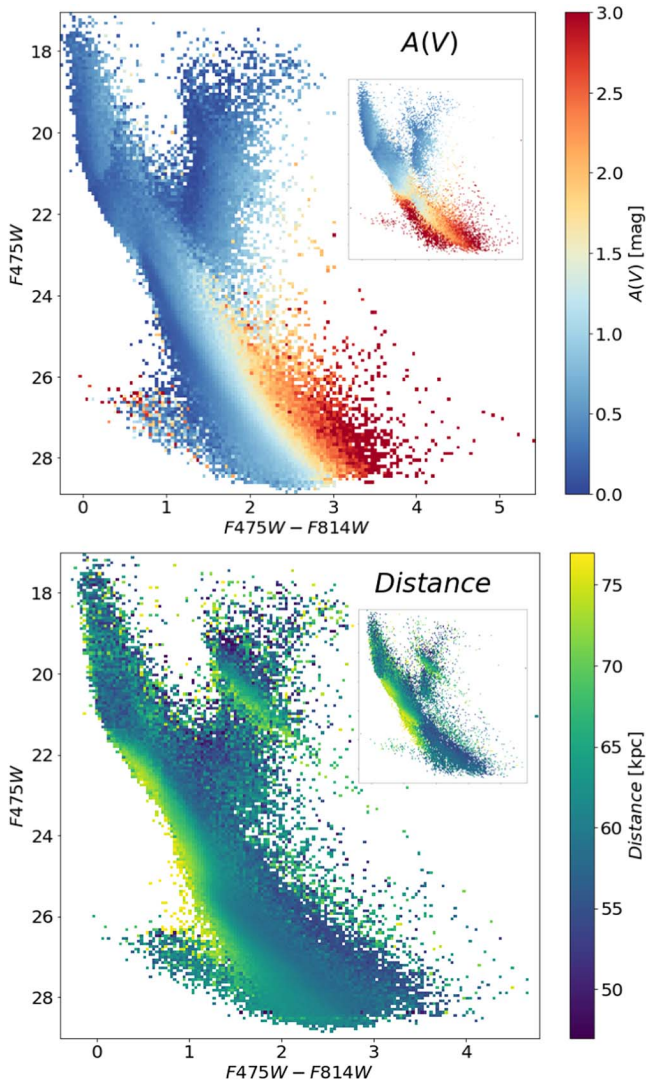

**Figure 7.** CMDs of BEAST fits color coded by the parameter indicated on each panel ($A(V)$ in the top panel and distance in the bottom panel). The colors correspond to the median value in each bin of the 2D histogram. The main CMDs are based on the `Basic Cuts` catalog while the insets show the `High SNR A(V)` catalog and show both the unreddened and reddened stellar populations in the SMIDGE field. The elongation of the RC and the widening of the RGB due to dust are notable.

independent CMD fitting method. This is an interesting result that alludes to the fact that the SMC is highly elongated along the line of sight (S. Subramanian & A. Subramaniam 2009; R. Haschke et al. 2012; A. M. Jacyszyn-Dobrzeniecka et al. 2016, 2017), most likely due to its dynamical interaction history with the LMC and the MW (G. Besla et al. 2007, 2012, 2016). As discussed in YMJ17 and YMJ21, these results have implications for measurements of dust extinction properties from the CMD.

### 4.4. Dust Extinction Properties

The 1D histograms in Figure 5 illustrate dust extinction features particular to the SMIDGE field. We show distributions for several thresholds in the $A(V)$ SNR to understand how quality cuts affect our results.

For the `High SNR A(V)` sample, we see an $R(V)$ distribution with a peak between 2.5 and 3, and also note a peak at $R(V) \sim 3.2$. The mean of the $R(V)$ distribution for the `High SNR A(V)` sample is 3.13. Although the main peak may be the result of consistency with our chosen $R(V)$ prior of 2.74, which is based on our best knowledge of SMC dust properties from observations, the secondary $R(V) \sim 3.2$ peak could be attributed either to the contribution of foreground MW $R(V) \sim 3.1$ dust, or to MW-type dust inherent in the SMC, or to both. We discuss these peaks and the implications for the SMC 3D structure in Section 5.1.2. It is of note that, based on an expanded sample of sight lines, the SMC $R(V)$ average has been recently updated from the G03 value of 2.74 to $R(V) = 3.02$ (K. D. Gordon et al. 2024), aligning more closely with our mean $R(V) = 3.13$).

Further, we see an $f_\mathcal{A}$ distribution peaked at $\sim 0.6$ (with a mean of 0.65), although we choose an $f_\mathcal{A}$ prior peaked at 1. This result may indicate that, in the SMC, $f_\mathcal{A}$ is on average lower than 1, while if it were prior-dominated, $f_\mathcal{A}$ would tend to 1. Additionally, we see a considerable fraction of stars with $f_\mathcal{A} \sim 1$ ($\sim 7\%$ of stars have $f_\mathcal{A} \geqslant 0.99$ in the `High SNR A(V)` catalog). Similarly to our interpretation of the secondary $R(V) \sim 3.2$ peak, we interpret this result as either a foreground contribution from diffuse ($R(V) = 3.1$) MW-type dust ($f_\mathcal{A} = 1$), or a contribution from SMC stars with MW-type dust, or both. It is of interest to explore these sight lines since, to date, there are only a handful of SMC sight lines that show a strong 2175 Å bump in their extinction curves (K. D. Gordon et al. 2024), while the majority lack a bump. These measurements may indicate that SMC dust grains have different properties from the average diffuse MW dust properties. We discuss additional fitting uncertainties in Section 4.5.

We note that the MW foreground toward the SMC ($A_V \sim 0.18$ mag; E. Muller et al. 2003; D. E. Welty et al. 2012) affects the full SMIDGE $A(V)$ distribution equally. In the `Basic Cuts` catalog, we see a double-peaked $A(V)$ distribution, with the first peak at a similar value for $A(V)$ ($\sim 0.18$ mag), and the second peak at $A(V) \sim 3$ mag. We speculate that this first peak is caused by extinction only due to the MW foreground, while the second peak is caused by the dust column intrinsic to the SMC *and* the foreground extinction. We therefore conclude, as in YMJ21, that distinct groups of stars cause the bimodality in $A(V)$, where one population is affected only by foreground dust (and not by dust intrinsic to the SMC), and another in addition experiences dust extinction due to dust intrinsic to the SMC. The `High SNR A(V)` catalog shows a strong single peak at $A(V) \sim 0.6$ mag (low-extinction sight lines are by definition excluded), and also a second peak at $A(V) \sim 2.35$ mag, which is subtle in the `High SNR A(V)` sample, but becomes more prominent and shifts to higher $A(V)$ values in the samples with higher S/N in $A(V)$ (see Figure 5). Overall, the results for $A(V)$ and the fact that it is clearly independent of its prior give us confidence that our dust extinction fitting is sensitive to the physical conditions of the dust column in the SMC.

The star in the individual BEAST-fit example in Figure 3 is also a part of the MR12 sample of stars (star 11) observed with HST/STIS UV prism spectra and optical/NIR photometry. This is not the most common type of star in the SMIDGE field as it is a reddened hot MS star for which MR12 measure a strong 2175 Å bump. For the rest of the stars, MR12 measure either a weak bump (star 08), or no bump (stars 09 and 10). K. D. Gordon et al. (2024) measure a significant bump for MR12-10 and MR12-11, and a weak or an absent bump for MR12-09 (they do not make a measurement of MR12-08). The BEAST fits for these stars are shown in Table 4. Our results for stars 08, 09, 10, and 11 are $f_\mathcal{A} \sim$ 0.99, 0.47, 0.31, and 0.45; thus, we find star 08 to have a high probability of a bump, but find a weak bump probability for star 10. For stars 09 and 11, the dust mixture coefficient indicates ~50% MW-type dust, or a medium-strength bump. Our fits show star MR12-08 to be significantly closer along the line of sight (at ~55 kpc) than the rest of the SMC B1 MR12 stars (at ~64–68 kpc), which may imply that this sight line is not a part of SMC B1-1, but rather is spatially coincident with it. In Section 5, we discuss the implications of finding stars with a high $f_\mathcal{A}$, or MW-type dust, in the SMIDGE field when most UV spectroscopy studies point to very few sources with a measured 2175 Å bump, and discuss how our catalog can be used to systematically identify all such sources.

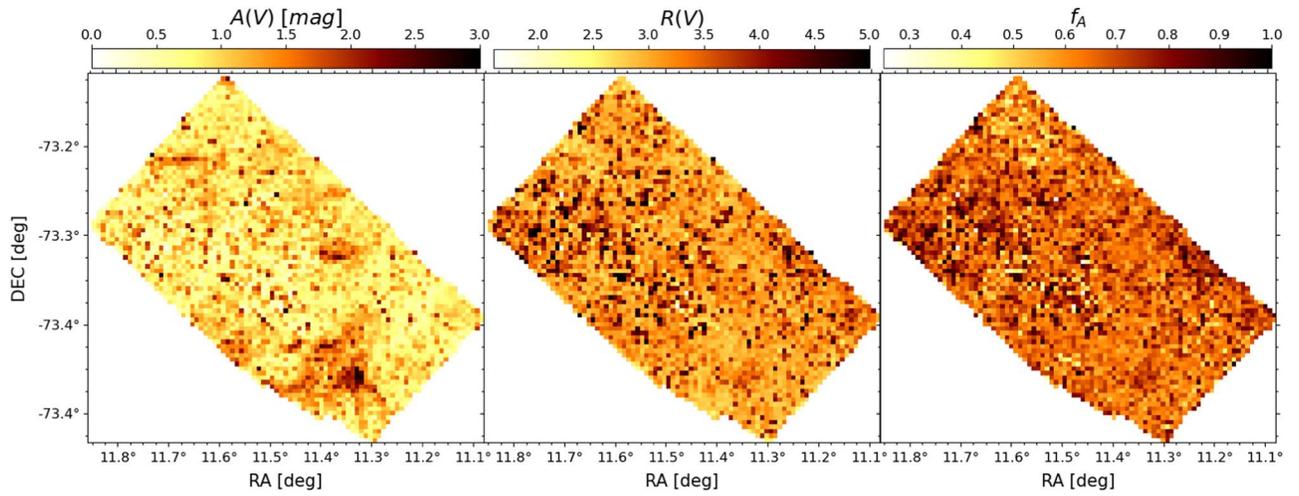

**Figure 8.** $A(V)$, $R(V)$, and $f_\mathcal{A}$ maps averaged to a 7″ resolution based on the BEAST-derived `High SNR A(V)` catalog. Left: higher $A(V)$ regions are noticeable in the SMIDGE footprint in several locations. Middle and right: the $R(V)$ and $f_\mathcal{A}$ maps at the same resolution are weighted by $A(V)$ to show these quantities for the total column of dust in a pixel. These two maps are relatively featureless, showing only a moderate increase toward the same molecular cloud regions.

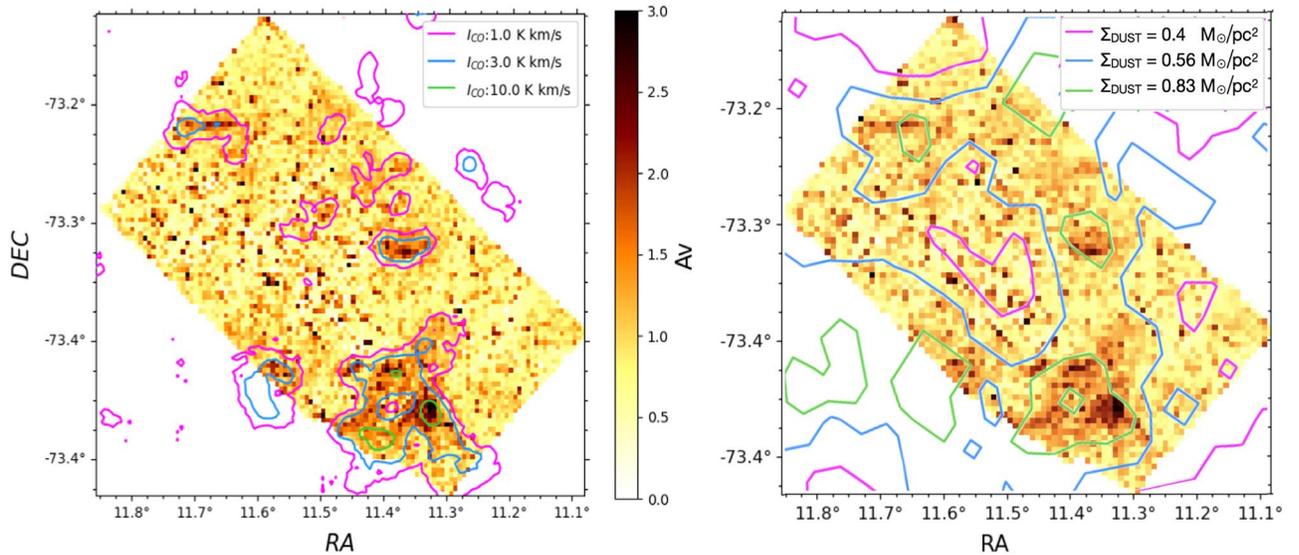

**Figure 9.** (Both panels) The SMIDGE `High SNR A(V)` dust map averaged to a 7″ resolution with a matching $A(V)$ color scale. Left: the contours of the APEX $^{12}$CO (2–1) emission map from H. P. Saldaño et al. (2023) are overlaid and illustrate the coincidence of high $A(V)$ and high $^{12}$CO regions. Right: contours of $\Sigma_{\rm dust}$ derived from IR emission at ~36″ (J. Chastenet et al. 2019) are overlaid. (The resolution of the latter is kept at 7″ for consistency with the left panel.) The visually apparent correlations between $A(V)$ and $I_{\rm CO}$, and also between $A(V)$ and $\Sigma_{\rm dust}$, are quantified in Figure 12.

One of the main questions we would like to answer about the SMC's dust properties, including dust extinction, is how the ISM environment may affect these properties. In Figure 8, we show maps of the BEAST-derived $A(V)$, $R(V)$, and $f_\mathcal{A}$ dust extinction parameters based on the median of the `High SNR A(V)` individual fits within a pixel of size 7″. The $R(V)$ and $f_\mathcal{A}$ maps are weighted by the amount of dust, $A(V)$, and they thus determine $R(V)$ and $f_\mathcal{A}$ of the total column of dust per pixel. Since the maps are based on the catalog with highly reliable $A(V)$ results, which in effect removes low $A(V)$ values (i.e., $A(V) \lesssim \sim 0.2$ mag), we do not expect a significant contribution from unreddened SMC stars, which only experience the MW foreground.

In Figure 9, we compare our $A(V)$ map to maps of ancillary ISM data, such as CO emission and the dust mass surface density ($\Sigma_{\rm dust}$) measured via fits to the IR emission from dust. Visually, we see a very good correlation between $A(V)$ and the

$^{12}$CO (2–1) emission (at 7″ resolution), and a slightly weaker correlation with $\Sigma_{\rm dust}$, where the resolution difference may account for the lack of spatial correlation. We also compare dust extinction fits with the H I emission tracing the distribution of neutral atomic ISM, and also to $q_{\rm PAH}$ maps, but we see no correlation with either. We explore the reasons behind these trends in Section 5.1.1.

### 4.5. Fitting Results Caveats

There are limitations to our fitting due to several factors. One limitation stems from the fact that, currently, the BEAST does not have models for binary stars or for stars with circumstellar dust, such as AGB stars. We do not expect, however, there to be a significant, if any, contribution from dusty, thermally pulsing AGB stars in the SMIDGE sample since these stars are bright and are all saturated: they are brighter than the tip of the RGB (TRGB; A. Bressan et al. 2012; P. Marigo et al. 2013), and the TRGB at the SMIDGE distance modulus (18.91–18.96; R. W. Hilditch et al. 2005; V. Scowcroft et al. 2016) would be at F475W ∼ 17.9 mag (there are only 76 stars at or brighter than this limit in the `High SNR A(V)` sample). As for early AGB stars, which are warmer and bluer than RGB stars, though their presence may cause a degeneracy with RGB stars, such stars are so short lived and rare that we do not expect them to contribute in a statistically significant way to extinction results for the RGB.

The BEAST code is in active development, and upon subsequent fitting we aim to implement models for such stars. Still, we are able to fit a variety of populations of stars with little circumstellar dust. It is important to note that, due to the wide wavelength coverage of the SMIDGE (and other multi-band) observations, the BEAST is distinct from other SED fitters (i.e., E. da Cunha et al. 2008; Y. Han & Z. Han 2014) in that it also fits the stellar luminosity, in addition to the flux. Fitting only for the latter, in the presence of dust extinction, would inevitably create degeneracies between, for example, MS, giant, and supergiant stars. By also fitting for the stellar luminosity *and* the stellar distance, we are able to derive intrinsic stellar parameters, which are fundamental to help distinguish between different stellar populations, and to break degeneracies between $\log(T_{\rm eff})$ and $A(V)$.

We also note that the precision of the recovery of the secondary parameters (which influence the SED more subtly), such as $R(V)$, $f_\mathcal{A}$, and $Z$, is lower than the precision of the recovery of the primary parameters, such as $A(V)$, $\log(t)$, and $\log(M)$, which are more sensitive to small changes in the SED shape and therefore drive the overall SED shape. For a more detailed discussion of BEAST parameter sensitivities, see Section 5.1 of G16. For example, we can see that, while the $A(V)$ results appear to be independent of the prior, $R(V)$ and $f_\mathcal{A}$ are not independent of the 2D $R(V)$ versus $f_\mathcal{A}$ prior in our dust mixture model. Still, we see features in the distributions of both parameters, such as secondary peaks at $R(V) \sim 3.2$ and $f_\mathcal{A} \sim 0.65$ (also evident in Figure 10, which we cannot explain simply with replicating the priors).

Specifically for SMIDGE, we illustrate the uncertainty in the dust extinction parameters fits in Figure 11. The precision of the recovery of the primary parameter $A(V)$ is overall very good, with a gradient clearly seen from bright to faint sources, pointing to the fact that $A(V)$ recovery is tied to the S/N in the observations (hence our decision to base our results on a `High SNR A(V)` subsample). The other primary parameters are recovered with a similar high precision. We note that $A(V)$ for upper- to mid-MS and RGB/RC stars is recovered very well with an average $1\sigma$ uncertainty of 0.12 mag. We posit that a fraction of the faintest stars with a F475W − F814W ∼ 2 mag and F475W ∼ 28 mag may be background sources that evaded our quality cuts.

The precision of the recovery of the secondary parameters $R(V)$ and $f_\mathcal{A}$ is lower and is inherent in the fact that observational noise affects these parameters more strongly. For example, the average $1\sigma$ uncertainty in $R(V)$ is 0.4, and the average $1\sigma$ uncertainty in $f_\mathcal{A}$ is 0.26. The $R(V)$ sources with the highest uncertainty are in a similar part of the CMD as those with higher uncertainty in $A(V)$, and for $f_\mathcal{A}$, sources throughout the lower part of the CMD have the highest uncertainty. This is similarly caused by the lower S/N for those faint sources. Additionally, as seen in Figure 4, there is an inherent difficulty in acquiring high precision on these two parameters with the SMIDGE filter set, which extends to the NUV F225W, while it is in the FUV part of the SED that we can obtain higher precision and be more sensitive to variations in these two parameters.

## 5. Discussion

### 5.1. Dust Extinction

#### 5.1.1. Maps

The BEAST fits to the SMIDGE data set produce the first SMC high-resolution dust column density maps at 7″ along hundreds of thousands of individual lines of sight (Figures 8 and 9). There is a moderate amount of dust in the SMIDGE field resulting in an $A(V)$ peak at ∼0.6 mag for the `High SNR A(V)` catalog (Figure 5 (mean ∼1.1 ± 0.83 mag)).

In Figures 8 and 9, which utilize the `High SNR A(V)` catalog, we see various degrees of correlation between the BEAST-averaged dust maps and ISM tracers in the SMIDGE field. We expect that dust will be well mixed with both molecular and atomic gas; thus, it should trace both types of gas equally well. But since we expect the dust column density to be higher toward molecular regions where there is more dust, on the one hand this should make dust easier to detect, while on the other we expect we are not able to measure all sight lines due to completeness. Particularly for the `High SNR A(V)` catalog, which is biased toward large $A(V)$ values (as seen in the top-left panel of Figure 5), this may eliminate lines of sight where gas is primarily atomic H I and $A(V)$ is low. We explore these correlations also using the `Basic Cuts` catalog, which contains ∼10 times more sources, and discuss these trends below.

Quantitative comparisons between the dust maps properties and ISM tracers are shown in Figure 12. We compare dust extinction parameter averages of sight lines found within bins of intensity in ancillary data maps (as described in the caption of the figure). We do not convolve our dust extinction maps to the resolution of the ISM maps, but instead take all individual fits within an ISM interval in aggregate. The $^{12}$CO (2–1) map at 7″ resolution observed with APEX (H. P. Saldaño et al. 2023) provides the finest spatial resolution of the ISM tracers we use. For CO, we observe the strongest correlation with the dust column $A(V)$, as seen in the upper middle panels of Figure 12 (Pearson $r = 0.89$). This trend can be interpreted as an indication that $A(V)$ is a relatively good tracer of CO, and vice versa. More fundamentally, C. Lee et al. (2015) note that, in the Magellanic Clouds, the correlation is expected since there cannot be CO emission without a substantial amount of dust extinction $A(V)$ (>1–2 mag).

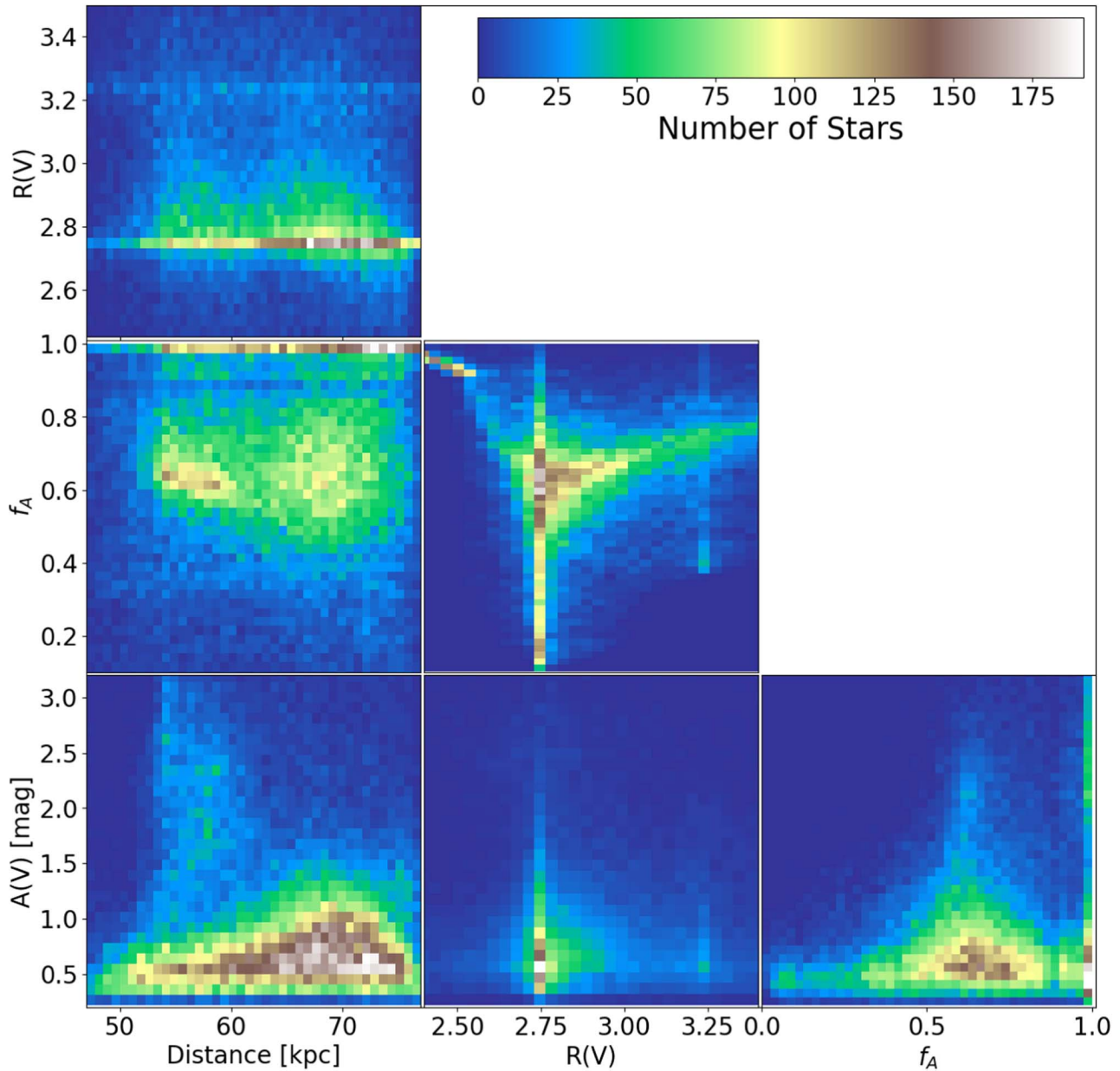

**Figure 10.** Two-dimensional correlation plots of the SMIDGE dust extinction parameters for the `High SNR A(V)` catalog. The color bar shows the density of points. We note several features: in the $A(V)$ vs. distance plot, there is a distinct high-$A(V)$ population at distances of 54–60 kpc, and a second distribution containing the majority of sources with lower $A(V)$ values. We similarly observe two distinct populations in the $R(V)$ vs. distance and $f_\mathcal{A}$ vs. distance plots. The $R(V)$ prior and the $f_\mathcal{A}$ prior peaking at 2.74 and at ∼1 are also seen (discussed in Section 4.5).

As for correlations between CO and $R(V)$, and between CO and $f_\mathcal{A}$, we do not observe a correlation with $f_\mathcal{A}$, while we see a weakly inverse correlation with $R(V)$ ($r = -0.49$). Additionally, while for $0.25 \leqslant I(CO) \leqslant 9.5\,\mathrm{K\,km\,s^{-1}}$ the correlation with $R(V)$ is weakly negative, for CO intensities above ∼9.5 K km s$^{-1}$, we observe a positive correlation with both $R(V)$ and $f_\mathcal{A}$ ($r = 1$ and 0.65, respectively). These trends may be related to extinction curve measurements in the SMC, where it is known that in SMC B1-1, for example, on very small scales, we see sight lines with both MW- and SMC-like dust properties, with and without a 2175 Å bump (MR12). If the positive correlation with $R(V)$ in dense CO regions is physical despite the low-number statistics, this may indicate that, in the SMC, there may be thresholds of cloud densities where dust becomes either progressively more SMC-like (a steeper UV rise and no 2175 Å bump) or more MW-like (diffuse ISM $R(V) \sim 3.1$, with 2175 Å bump). One physical interpretation may be that, at a certain CO threshold, small dust grains (lower $R(V)$) begin to coagulate and form larger dust grains (higher $R(V)$). These results are consistent with the results of J. B. Foster et al. (2013) for MW dust in the Perseus molecular cloud, who found that, beyond $A(V)$ of 2 mag, there is a positive correlation between $R(V)$ and $A(V)$, indicating a change from diffuse to dense ISM (although in our study the average value of $A(V)$ per CO intensity bin does not surpass 2 mag due to the coarse resolution of bins of CO intensity). Variations in $R(V)$ can also be seen in the rich data provided by 3D dust maps of the MW. E. F. Schlafly et al. (2017) and

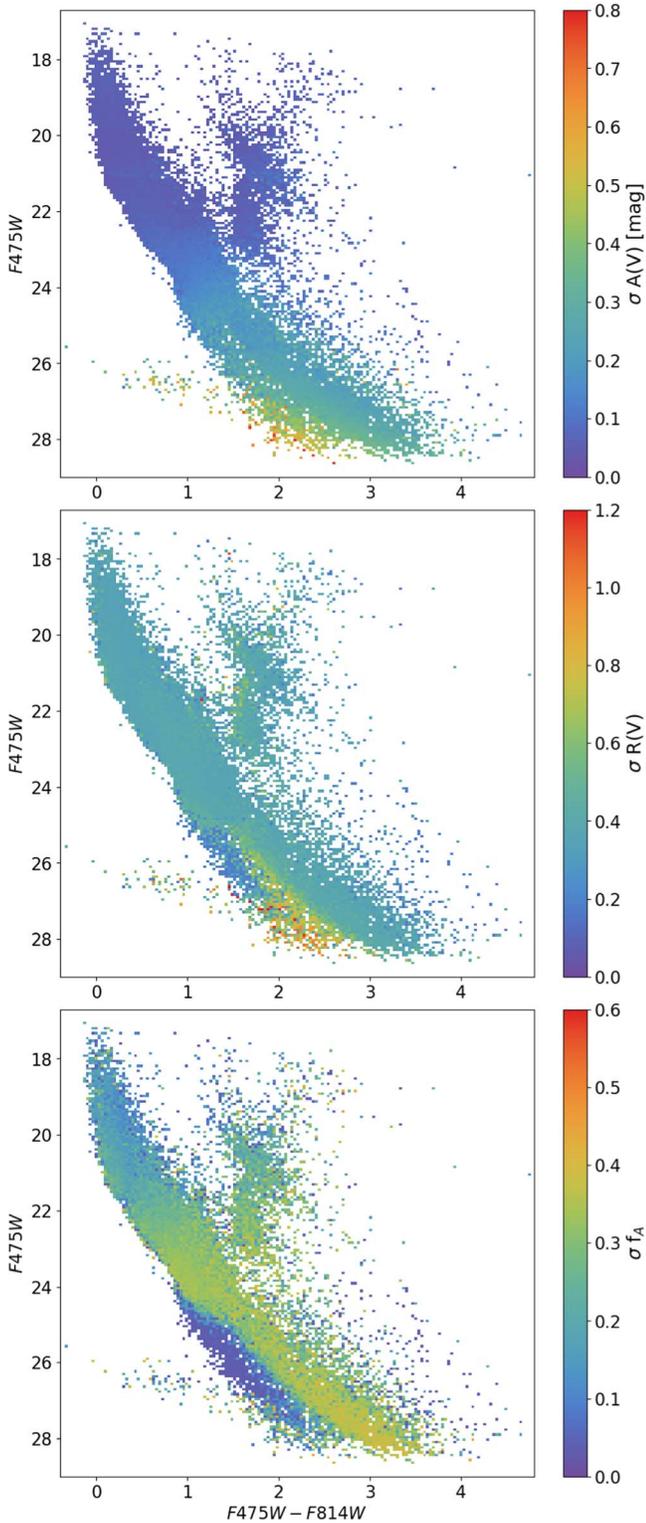

**Figure 11.** CMDs of BEAST fits color coded by the $1\sigma$ uncertainty in $A(V)$ (top), $R(V)$ (middle), and $f_\mathcal{A}$ (bottom). The colors correspond to the median value in each bin of the 2D histogram. The CMDs are based on the High SNR A(V) catalog.

G. M. Green et al. (2018), for example, generate such maps on large galactic scales. They observe that $R(V)$ variations are seen on kiloparsec scales, and do not see a strong connection between the large-scale structure of $R(V)$ variations in the outer Galaxy and grain growth in dense regions of the ISM.

Using the Basic Cuts catalog for this comparison, for CO we see a very similar trend for $A(V)$ ($r=0.91$), only the $A(V)$ range is much narrower and is between 0.6 and 1 mag. Further, we observe more pronounced negative correlations for $R(V)$ and $f_\mathcal{A}$ ($r=-0.88$ and $r=-0.8$, respectively). For $R(V)$, there is an indication of a cutoff $I$(CO) threshold of $\sim 8\,\mathrm{K\,km\,s^{-1}}$, where it appears that within denser ISM regions, $R(V)$ centers around the average diffuse MW value of $\sim 3.1$, while in the more diffuse ISM, $R(V)$ increases to $\sim 3.15$. Thus, while overall there is a negative correlation between CO and $R(V)$ in both the High SNR A(V) and the Basic Cuts catalogs, there is a difference in the average value $R(V)$ assumed within the same bins of CO-integrated intensity. As for $f_\mathcal{A}$, the Basic Cuts catalog results show a decreasing value with CO-integrated intensity, suggesting dust becomes more SMC- and less MW-like.

We also compare the dust mass surface density derived from IR emission to $A(V)$ on $\sim 36''$ per pixel scales using the J. Chastenet et al. (2019) map based on Spitzer and Herschel observations fit with the B. T. Draine & A. Li (2007) dust model. We find a positive correlation between $A(V)$ and $\Sigma_{\mathrm{dust}}$ (Pearson $r=0.63$). We also make a comparison on slightly larger spatial scales with the $\Sigma_{\mathrm{dust}}$ map of K. D. Gordon et al. (2014), and again find a correlation between $A(V)$ and $\Sigma_{\mathrm{dust}}$ ($r=0.68$). Using either $\Sigma_{\mathrm{dust}}$ map, however, we find no strong correlations between $\Sigma_{\mathrm{dust}}$ and $R(V)$ or $f_\mathcal{A}$.

We also use the $q_{\mathrm{PAH}}$ map at $\sim 36''$ of J. Chastenet et al. (2019) to explore correlations between PAHs and $f_\mathcal{A}$, and also between PAHs and $A(V)$. Since the latter indicates the probability for the presence of the 2175 Å bump in the extinction curve (see Section 3.2), we use this comparison to test whether PAHs may be the carriers of the bump. While here we are not comparing ancillary ISM tracers directly to each other, other SMC studies such as that of K. M. Sandstrom et al. (2010) do find a correlation between, say, $q_{\mathrm{PAH}}$ and the CO intensity, whose correlation we examine here. Additionally, using new HST/STIS and archival IUE UV spectroscopic measurements toward 25 SMC sight lines, K. D. Gordon et al. (2024) find a correlation between $q_{\mathrm{PAH}}$ and the 2175 Å bump area (height + width). We do not, however, find a strong, if any, correlation between $q_{\mathrm{PAH}}$ and $f_\mathcal{A}$, or between $q_{\mathrm{PAH}}$ and $A(V)$ or $R(V)$, therefore our results do not allow us to make a definitive statement about the relationship between PAHs and the bump.

We examine in more detail the apparent lack of correlation between $q_{\mathrm{PAH}}$ and $A(V)$. If we divide the $q_{\mathrm{PAH}}$ range into sections, we note that, while for low and high $q_{\mathrm{PAH}}$ values in the range $1.1 \leqslant q_{\mathrm{PAH}} \leqslant 1.7\%$ and $2.5 \leqslant q_{\mathrm{PAH}} \leqslant 2.8\%$ we do not find a correlation, for the midrange of $1.75 \leqslant q_{\mathrm{PAH}} \leqslant 2.45\%$, we find strong correlation with a Pearson $r=0.93$. While we ideally would look at the entirety of the $q_{\mathrm{PAH}}$ and $A(V)$ ranges within SMIDGE, these trends are of note, and merit further exploration.

We note that our catalog can be used to identify bump-candidate sight lines that may be suitable for spectroscopic follow-up. We identify $\sim 200$ sight lines in the High SNR A(V) sample within the SMIDGE footprint with $f_\mathcal{A} \geqslant 0.9$ that are relatively reddened and bright with $A(V) \geqslant 0.5$ and F225W $\leqslant 20$. We will also make this catalog available as an HLSP upon publication.[13]

We do not see a clear positive correlation between the neutral hydrogen emission in the SMC (N. M. Pingel et al.

---
[13] doi:10.17909/f9wx-e637.

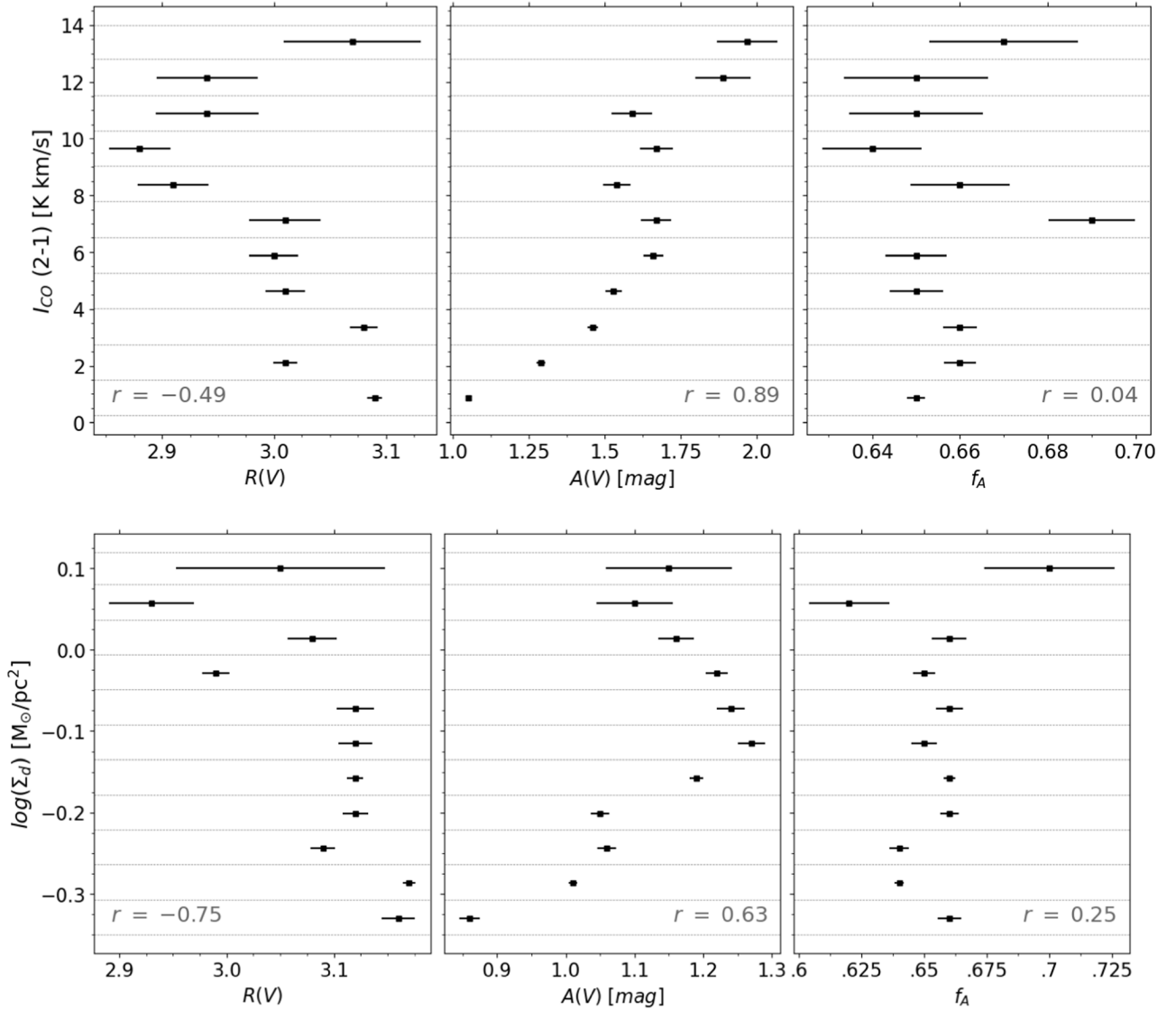

**Figure 12.** Correlations between two ISM tracers and SMIDGE SMC dust extinction properties determined for the `High SNR A(V)` catalog. The top panels show the correlation with the intensity of the $^{12}$CO (2–1) emission ($I_{CO}$), and the bottom panels show the correlation with the dust mass surface density, $\Sigma_{dust}$. Dashed lines indicate the interval range of each ISM tracer to which sight lines spatially coincident with this region of ISM intensity are compared. We see a positive correlation between $A(V)$ and $I_{CO}$, and also between $A(V)$ and $\Sigma_{dust}$ ($r = 0.89$ and $r = 0.63$, respectively). We see a negative correlation between the ISM tracers and $R(V)$ ($r = -0.49$ and $r = -0.75$, respectively). There is a lack of correlation with $f_A$. The $A(V)$ correlation indicates that these two tracers may also be used as tracers of the dust column density. As the values of both $I_{CO}$ and $\Sigma_{dust}$ increase, uncertainties on the dust parameters increase due to low-number statistics.

2022) and the dust parameters we fit, regardless of which catalog we use (`Basic Cuts` or `High SNR A(V)`) when we take into account the full range of $N$(H I) within the SMIDGE field ($\sim 8 \times 10^{21}$–$1 \times 10^{22}$ cm$^{-2}$). For $A(V)$, we find a Pearson $r = -0.12$ for the former and $r = -0.41$ for the latter sample. Comparing the correlations with $A(V)$ at higher $N$(H I) sight lines within the SMIDGE field ($\geqslant 9.1 \times 10^{21}$ cm$^{-2}$), we note that, for both samples, we see strong positive correlations of $r = 0.94$ and $r = 0.72$ for the `Basic Cuts` and the `High SNR A(V)` samples, respectively (where the sight lines that are spatially coincident with $N$(H I) beyond threshold comprise $\sim 25\%$ and $\sim 2.5\%$ of the total respective samples). Although generally we would expect that dusty sight lines trace both the molecular and atomic ISM, and vice versa, $N$(H I) in the SMIDGE field is somewhat low and spans only a small range; thus, a comparison in other SMC regions with higher $N$(H I), and also in the LMC, would be more informative.

### 5.1.2. Two-dimensional Dust Correlations

In Figure 10, we show the 2D correlations among the dust extinction parameters and the distance. The dust extinction–distance distributions can inform our understanding of the 3D geometry of the SMC SW bar, including where the dust lies relative to the stars. Several features of our results are noteworthy. For example, we see two structures with distinct average distances of $\sim 57$ and $\sim 68$ kpc, corresponding to the two peaks in the distance distribution in Figure 5. A recent analysis by C. E. Murray et al. (2024a) of the 3D SMC ISM structure using neutral hydrogen measurements also finds that the galaxy is composed of two distinct components separated by $\sim 5$ kpc along the line of sight, at average distances of 61 and 66 kpc. Similarly, YMJ21 conclude that the SMIDGE field is in a region containing a thin dust layer located at $\sim 62$ kpc, which is in front of the bulk of the stellar population (resulting in a reddened fraction of stars greater than 50%). It is remarkable that these three independent methods of constraining the SMC 3D structure are consistent with each other.

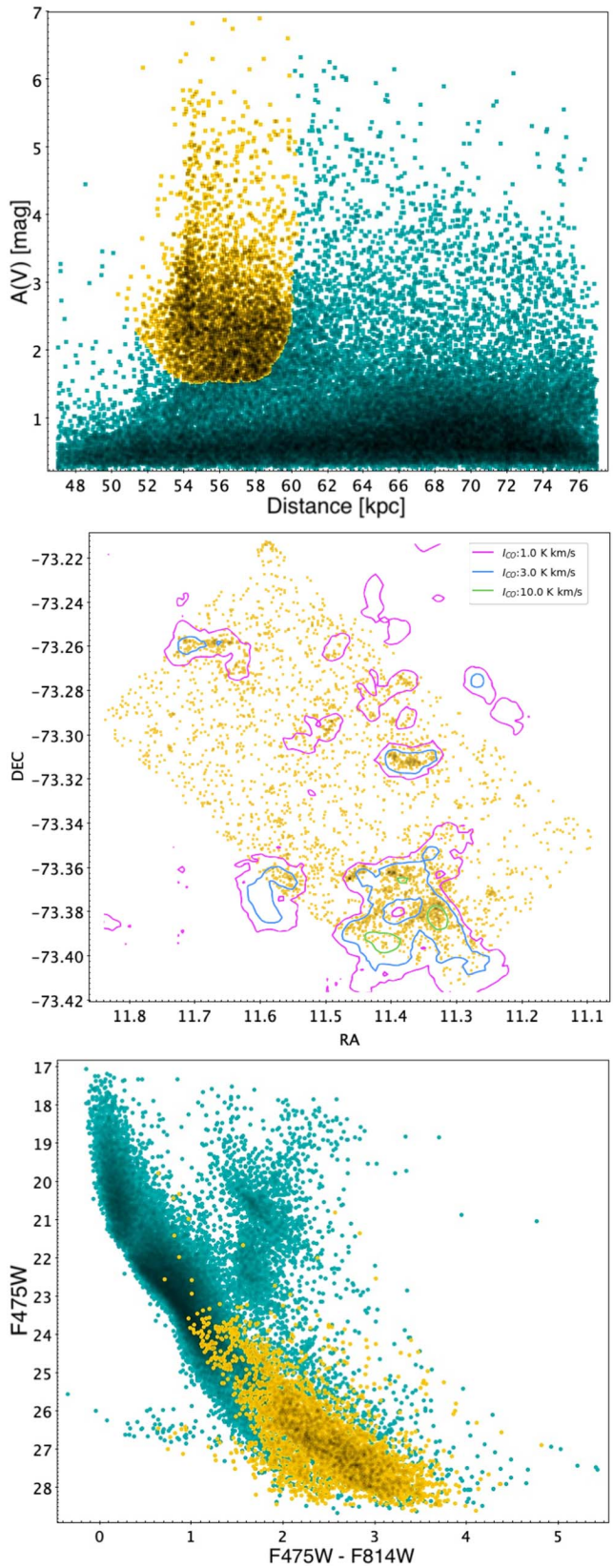

**Figure 13.** We explore the nature of the prominent population of high-$A(V)$ stars at distances slightly closer than the centroid of the SMC stellar population (62 kpc; upper panel, yellow points). The location of many of these sources seem to be spatially coincident with high-CO regions, as indicated by the CO intensity contours (same as in Figure 9). The majority of these stars are composed of lower MS and PMS stars.

We explore in more detail the distinct high-$A(V)$ population associated with the stellar component at ∼55–60 kpc seen in Figure 10, which resides in front of the average SMC distance (62 kpc; V. Scowcroft et al. 2016). This population is the cause of the secondary $A(V)$ peak seen in Figure 5. We analyze this population in Figure 13 by placing a cut on $A(V) \geqslant 1.7$ mag and a range on distances of 52–59 kpc. In the middle panel, we note the spatial coincidence between these nearside sight lines and molecular gas traced by CO emission. (About 13% of these sight lines are located within regions with $I(CO) \geqslant 1.0$ K km s$^{-1}$ (see Figure 9).) These results are consistent with C. E. Murray et al. (2024a), who show that the CO emission in the SMC is associated with the front ISM component in their study. As for the nature of the stellar sources, we observe in the lower panel of the figure that the vast majority of these stars are low-MS and PMS sight lines. It is notable that the sample does not include many upper MS and RC or RGB stars, where we used the latter in YMJ21 in CMD fitting to obtain dust extinction and distance results for the SMC. We speculate that this is based on the twofold effect, where these stellar populations either reside in sight lines with $A(V) \leqslant 2$ mag, or the sight lines with higher $A(V)$ are dominated by PMS stars. An inverse test (selecting stars in the PMS region of the CMD) shows an almost exclusive spatial association with high-CO regions, and gives us confidence in our fits, confirming PMS stars begin their lives in dense molecular cloud regions. One reason we do not see the high-$A(V)$ component associated with young, massive stars may be due to the limited completeness of our observations for such stars (as such stars would most certainly also form in molecular cloud regions).

## 6. Conclusions and Further Work

We obtain the first catalog of individual stellar and dust properties based on deep HST photometric observations of ∼20 kpc$^2$ in the SW bar of the SMC. We use the BEAST to fit seven-band HST data to constrain these properties for a great variety of stellar types, ranging from blue supergiants to red dwarfs. The BEAST allows us to disentangle potentially degenerate effects on the SED shape, i.e., those due to temperature and extinction, encompassing $A(V)$, $f_A$, and $R(V)$. We also present maps of stellar and dust properties compared to the resolution of existing ISM data for a holistic view of the interaction of the stars, gas, and dust in the SMC.

Our dust maps show correlations between dust extinction and ISM tracers. Notably, there is a strong correlation between dust extinction and CO emission. This result is consistent with findings in the Magellanic Clouds that, to have CO emission, there needs to be a significant amount of dust extinction. Additionally, we explore correlations between dust parameters and the presence of the 2175 Å bump in the extinction curve, as well as correlations between dust mass surface density and other dust parameters.

Further analysis reveals distinct structures in the dust extinction–distance distributions, providing insights into the 3D geometry of the SMC. These findings are consistent with previous studies and suggest a multicomponent structure for the SMC. Additionally, we examine the association of high dust extinction regions with molecular gas and PMS stars, shedding light on the connection between star formation and dust distribution. Overall, our work provides valuable insights into the stellar properties and dust extinction in the SMC, contributing to our understanding of star formation processes and interstellar dust in galaxies.

Regarding stellar properties, our study reveals peaks in the stellar age distribution that may indicate periods of enhanced star formation. These findings align with previous studies of the star formation history of the SMC. The distribution of stellar masses also shows interesting features, with a significant peak around $1\,M_\odot$ and an enhancement around $5\,M_\odot$.

We leave an analysis of the dust-to-gas ratio $A(V)/N$(H) for subsequent work (P. Yanchulova Merica-Jones 2024, in preparation), which will explore this and other dust extinction properties in both the SMC and the LMC, also using data from the Scylla pure-parallel HST survey (C. E. Murray et al. 2024b).

## Acknowledgments

We would like to acknowledge that this work was made possible in part by NASA through grant Nos. HST GO-15891, GO-16235, GO-16786, and GO-13659 from the Space Telescope Science Institute, which is operated by AURA, Inc., under NASA contract NAS 5-26555. Some of the data presented in this article were obtained from the Mikulski Archive for Space Telescopes (MAST) at the Space Telescope Science Institute. The specific observations analyzed can be accessed via doi:10.17909/p5c8-yw20. Additionally, high-performance supercomputer work was supported by the National Science Foundation through the use of the Advanced Cyberinfrastructure Coordination Ecosystem: Services & Support supercomputer program (ACCESS; grant # PHY220054). These observations are made with the NASA/ESA Hubble Space Telescope and are associated with program # GO-13659. We made extensive use of NASA's Astrophysics Data System bibliographic services. This research made extensive use of the BEAST (G16), Astropy, a community-developed core Python package for Astronomy (Astropy Collaboration et al. 2013), NumPy (S. van der Walt et al. 2011), and Matplotlib (J. D. Hunter 2007). We also acknowledge the Aspen Center for Physics, supported in part by the National Science Foundation, which provided opportunities for discussing the work and ideas presented in this paper.

We thank multiple people for illuminating conversations and feedback on various parts of the manuscript. In particular, the authors would like to thank Martha Boyer, Christopher Clark, Marjorie Decleir, and Dries van de Putte. We would also like to thank the anonymous referee who helped expand the discussion in a number of illuminating ways.

## ORCID iDs

Petia Yanchulova Merica-Jones https://orcid.org/0000-0002-9912-6046
Karl Gordon https://orcid.org/0000-0001-5340-6774
Karin Sandstrom https://orcid.org/0000-0002-4378-8534
Claire E. Murray https://orcid.org/0000-0002-7743-8129
L. Clifton Johnson https://orcid.org/0000-0001-6421-0953
Julianne J. Dalcanton https://orcid.org/0000-0002-1264-2006
Julia Roman-Duval https://orcid.org/0000-0001-6326-7069
Jeremy Chastenet https://orcid.org/0000-0002-5235-5589
Benjamin F. Williams https://orcid.org/0000-0002-7502-0597
Daniel R. Weisz https://orcid.org/0000-0002-6442-6030
Andrew E. Dolphin https://orcid.org/0000-0001-8416-4093